\documentclass[12pt]{article}

\usepackage{times}

\usepackage{graphicx}
\usepackage{fixltx2e} 

\usepackage{fullpage}  

\usepackage{array}
\usepackage{longtable}
\usepackage[width=0.9\textwidth]{caption}

\newcommand{\mybox}[1]{\begin{center}
    \fbox{\parbox{0.94\textwidth}{#1}}
  \end{center}}

\title{Using Students as Experimental Subjects in Software
    Engineering Research --- A Review and Discussion of the Evidence}
\date{}

\author{Dror G. Feitelson\\
  School of Computer Science and Engineering\\
  The Hebrew University, 91904 Jerusalem, Israel}

\renewcommand{\dbltopfraction}{.8}
\renewcommand{\dblfloatpagefraction}{.801}
\newcommand{\hide}[1]{}

\newcommand{\code}[1]{\textsf{\small #1}}

\begin{document}
\widowpenalty=5000 \displaywidowpenalty=5000 \clubpenalty=5000
\predisplaypenalty=50 \postdisplaypenalty=50

\maketitle

\begin{abstract}
  Should students be used as experimental subjects in software
  engineering?
  Given that students are in many cases readily available and cheap
  it is no surprise that the vast majority of controlled
  experiments in software engineering use them.
  But they can be argued to constitute a convenience sample that may
  not represent the target population (typically ``real''
  developers), especially in terms of experience and proficiency.
  This causes many researchers (and reviewers) to have reservations
  about the external validity of student-based experiments, and claim
  that students should not be used.
  Based on an extensive review of published works that have compared
  students to professionals, we find that picking on ``students'' is
  counterproductive for two main reasons.
  First, classifying experimental subjects by their status is merely a
  proxy for more important and meaningful classifications, such
  as classifying them according to their abilities, and effort should
  be invested in defining and using these more meaningful
  classifications.
  Second, in many cases using students is perfectly reasonable, and
  student subjects can be used to obtain reliable results and further
  the research goals.
  In particular, this appears to be the case when the study involves
  basic programming and comprehension skills, when tools or
  methodologies that do not require an extensive learning curve are
  being compared, and in the initial formative stages of large
  industrial research initiatives --- in other words, in many of the
  cases that are suitable for controlled experiments of limited
  scope.\\[2mm]
\textbf{keywords:} Controlled experiment; Experimental subjects;
Students; Experimental methodology.
\end{abstract}

\section{Introduction}

Experimentation, and especially experiments involving human subjects,
are not widely used in computer science \cite{tichy95,feitelson:exp}.
But two sub-fields are an exception:
human-computer interaction and software engineering.
In these disciplines the recognition that humans are inherently ``in
the loop'' has led to significant experimental work involving humans.
In particular, in software engineering controlled experiments in the
lab are used to learn about the effectiveness of different
methodologies, procedures, and tools \cite{basili86,basili07}.

However, experimenting with humans is difficult.
One of the salient problems is finding subjects who are
representative of the target population.
While a specific target population is often left undefined, in
software engineering it is typically understood to be the community of
software developers, and especially those who do it for a living.
But recruiting professional developers for experiments is hard,
because they need to be paid a competitive fee \cite{basili96c},
and there may be additional constraints such as not
interfering with normal production work \cite{dieste13}.

An intriguing alternative is to use computer science or software
engineering students, who are naturally much more accessible in an
academic setting, and in addition may be argued to constitute the next
generation of software professionals \cite{kitchenham02,tichy00}.
Experiments can sometimes even be done as part of class requirements
(e.g.\ \cite{juristo03,briand05,arisholm06,walia13}).
And indeed, a literature study of the period 1993--2002 found that in
87\% of the experiments reported students were used as subjects
\cite{sjoberg05}.
But this leads to the question of whether the results of student-based
experiments have external validity and generalize to real-life settings.
A similar situation occurs in psychology, where a majority of studies
are based on students who are WEIRD (Western, Educated,
Industrialized, Rich, and Democratic) and may not reflect
general human behavior \cite{henrich10}.

The problem of what experimental subjects to use has been recognized
long ago.
In one of the first papers to report on a controlled experiment in
software engineering, Myers writes concerning
``the question of whether one can extrapolate, to a typical industrial
environment, experimental results obtained from trainee programmers or
programmers with only a few years of experience'' \cite{myers78}.
Some time later, Curtis provocatively entitled a review of empirical
software engineering ``By the way, did anyone study any real
programmers?'' \cite{curtis86}.
More recently, Di Penta et al.\ write in the context of advising on
experimental procedures that ``a subject group made up entirely of
students might not adequately represent the intended user population''
\cite{dipenta07}, and Ko et al., in the context of discussing
experimentation for evaluating software engineering tools, write that
``undergraduates, and many graduate students, are typically too
inexperienced to be representative of a tool's intended user''
\cite{ko15}.
Jedlitschka and Pfahl, in suggested reporting guidelines for empirical
research, use ``Generalization of results is limited due to the fact
that undergraduate students participated in the study'' as an example
of a ``limitations'' clause that should be included in structured
abstracts \cite{jedlitschka05}.
Finally, in a literature survey focused on confounding factors in
program comprehension studies, the subjects' level of experience was
the most often cited factor \cite{siegmund15}.
This has led some researchers to actually hire professional
programmers to participate in extended experimental evaluations of
various techniques and tools
(e.g.\ \cite{sjoberg02,arisholm07,dzidek08,bergersen12,sjoberg13,yamashita13}).
But is this expenditure really necessary? And under what conditions?

Conversely, the obsession with ``students'' being a problem can be
criticized as reflecting a simplistic belief that graduation is a
pivotal event of great significance.
A more reasonable view is that there exists a wide range of
capabilities in both students and professionals, with large overlap,
but perhaps also focused improvements due to on-the-job training.
Tichy, for example, in a list of suggestions to potential reviewers of
empirical software engineering research, writes that ``don't dismiss a
paper merely for using students as subjects'' \cite{tichy00}.
Among the reasons he cites are the observation that even if this is
not the optimal setting much can still be learned from student
studies, and that students are in fact quite close to the target
population of developers \cite{kitchenham02}.
Likewise, Soh et al.\ discuss the classification of experimental
subjects based on professional status (students vs.\ professionals) as
opposed to a classification based on expertise, and conclude that
expertise does not necessarily correlate with professional status
\cite{soh12}.
Consequently excluding students just because they are students would
be detrimental both to research and to staffing decisions.
And a literature study from 2000--2010 found that large numbers of
professional developers are in fact not needed for effective user
evaluations, and moderate numbers of students can also be used
\cite{buse11}.
More generally, Hannay and J{\o}rgensen explain how absolute realism
is not always the best approach in experimental research, and that
artificial constructs may be preferable in certain cases \cite{hannay08}.
Finally, one should note that the goal of software engineering
experiments using human subjects is usually the evaluation of tools,
procedures, or methodologies.
Such evaluations are typically relative, trying to assess whether one
approach is better than another.
Therefore the requirement that subjects be representative may be an
overstatement: it is actually enough that the relative results be
representative.

Our goal in this review is to discuss the basic question:
\begin{center}
  \emph{Should students be used as experimental subjects in software\\
    engineering research, and if so, under what conditions?} 
\end{center}
To do so we organize and present the cumulative experience with
issues related to using students in software engineering research,
as it is reflected in the scientific literature.
This is structured according to the following sub-questions:
\begin{itemize}\itemsep 0pt
\item \emph{Why are students perceived to be a problem?}
  Section \ref{sect:stud} presents a list of the potential problems
  that may arise when using students as subjects, such as the
  differences between students and professionals in tool
  use and experience.
\item \emph{What is the evidence?}
  Section \ref{sect:dir} reviews studies that contrast students and
  professionals head-on.
  It reports evidence which shows both that the distinction is not
  important, so students can indeed be used in lieu of professionals,
  but also evidence that experience actually does matter, and thus
  that student novices and professional veterans may have widely
  different skills and require (or at least benefit from) different
  tools and procedures.
\item \emph{What does it all mean?}
  Given the wide diversity of results in the literature, Section
  \ref{sect:disc} provides a discussion of their implications,
  couching it in the more general treatment of experience and
  expertise.
  The main conclusion is that the focus on the dichotomy between
  students and professionals is a dangerous over-simplification.
  First, this is the wrong distinction, and it is much more
  important to classify experimental subjects according to their
  abilities and suitability to the specific study at hand.
  At the same time, one must realize that there is an extremely wide
  spectrum of software developers, so there is no single ``right'' way
  to select experimental subjects who will be generally
  representative.
\item \emph{How should students be used?}
  Finally, Section \ref{sect:rec} lists some recommendations
  concerning the use of students in empirical studies, in case you
  decide to do so.
\end{itemize}
There have been many papers that purport to address the question of
using students, but in the end this is always done in the context of a
single experimental setup.
This review attempts to pool all these previous studies, which are
reviewed in detail in Section \ref{sect:dir}.

It should be noted that experimental subjects (and in
particular, using students) are not the only potential problem with
controlled experiments in software engineering.
Actually, there are two main contentious dimensions in such experiments:
the experimental subjects and the tasks being tackled by these
subjects (Figure \ref{fig:prob})
\cite{brooksre80,sjoberg03,sonnentag06,carver10}.
In many cases the tasks are just toy problems in order to make running
the experiments manageable.
Such toy problems cannot coax out deep knowledge on how real experts
approach complex situations.
Moreover, the environment is also important.
Perry et al.\ point out that professional programmers spend only about
half of their time coding, so studies of various methodologies and
tools conducted under sterile laboratory conditions may not represent
what happens in real life situations \cite{perry94}.
It therefore seems that there are significant issues that are simply
out of the reach of controlled laboratory experiments.
While this is mentioned in the discussion, most of the review focuses
on the issue of subjects in controlled experiments, as it is already
wide enough.

\begin{figure}\centering
  \begin{tabular}{l|l|l}
    			& \emph{students} & \emph{professionals} \\
    \hline
    \emph{toy problem}	& academic experiment & industrial experiment \\
    \hline
    \emph{real problem}	& project/internship  & real-life situation \\
  \end{tabular}
  \caption{\label{fig:prob}\sl
  Differences between controlled studies and real-life situations.}
\end{figure}

We note too that using students is but one aspect of potential
threats to external validity.
And the whole issue of the relative importance of external validity as
opposed to internal validity is also hotly debated \cite{siegmund15b}.
Of course, if one believes that internal validity trumps external
validity the concerns about using students are moot.
But given that many researchers believe that external validity is more
important, to the point of expressing complete disbelief in academic
studies based on students, a review of this topic is believed to be
useful.

The methodology followed is a classical literature review.
This is not a ``systematic review'' \cite{kitchenham09} as the goal
is not to amass evidence for or against a certain effect.
Rather, the goal is to collect all the differing considerations and
opinions that have been expressed on this topic.
To do this we started with several sources, including 
keyword searches such as ``students'' and ``experimental subjects''
and snowballing (following the references in papers that had already
been read).
This enabled the identification of other important publications across
more venues and years than would be possible to scan systematically,
as well as the inclusion of literature from related fields, such as
general studies of experts vs.\ novices.
The price paid is that others who perform a similar review might end
up with a somewhat different list of papers.
The following sections present what we learned from the papers that
were found.

\section{Problems with Students}
\label{sect:stud}

The main concern regarding using students as experimental subjects is
that they may not represent professionals faithfully.
Thus if we are interested in software engineering as it is practiced
by professionals in the field, data obtained from student experiments
may be misleading.

There are several reasons why professional experience is important.
One is that experience can be expected to hone the knowledge
acquired during studies, making it more directly usable in performing
software engineering tasks and eradicating misconceptions.
Another is that experience leads to both improved and new
capabilities, even changing the way professionals approach a task and
not only the proficiency with which they attack it
\cite{lord90,schenk98}.
In addition, students (and experiments based on students) may also
suffer from the characteristics of the academic setting.
In this section we review studies that have demonstrated and discussed
such effects.

\subsection{Learning Misconceptions}

As noted above the concern about using students in empirical software
engineering research is that they ``lack experience''.
One aspect of this is that they may not have ingested all that they
have learned, or worse, have not learned something at all or have
learned it wrong \cite{guzdial11}.

A major issue that has been identified in the literature is that
novice programmers may hold invalid or inappropriate mental models of
what programming constructs actually mean.
For example, Bayman and Mayer studied how well 30 undergraduate
college students understood nine basic BASIC statements after a
6-hour self-study course widely used in the microcomputer lab
\cite{bayman83}.
The results showed that between 3\% and 80\% (median 27\%) achieved a
fully correct conceptual model of the statements, with up to 56\%
having incomplete models and up to 60\% having incorrect models.
Importantly, all subjects successfully completed the course,
indicating that the used tests of how well students master the
material were ineffective in discerning these problems.
Ebrahimi conducted a similar study, classifying language construct
misunderstanding by novice programmers learning Pascal, C, Fortran,
and Lisp \cite{ebrahimi94}.
In addition, he also noted errors in composing these constructs into
problem solution plans.
And Kaczmarczyk et al.\ found basic misconceptions regarding the
relationship between language elements and underlying memory use, how
while loops operate, and the object concept \cite{kaczmarczyk10}.

Similar results have been obtained also when studying more advanced
concepts.
For example, Kahney performed a study of how novice programmers
differ from experienced ones in their understanding of recursion
\cite{kahney83}.
The results from 30 novice students and 9 more experienced ones (TAs)
was that over half of the novices appear to have thought of recursion
as a loop, and only 3 really understood it, as opposed to 8 of
the more experienced ones.

More recently, Ma et al.\ looked at how students understood reference
assignment in Java at the end of their first year \cite{ma07}.
They found that only 17\% had a consistently viable model of what is
going on.
The remaining 83\% were divided evenly between those who were
consistently wrong and those who were inconsistent, both groups
displaying a wide variety of inappropriate models.
In this case a strong correlation was found between the viability of
the model held by the students and their performance in the final
exam.
Nevertheless, many students with non-viable models also managed to
perform well in the finals, and some even held non-viable models of
the simpler value assignment statement.
Likewise, Madison and Gifford found that two students managed to
construct correct programs and answer questions despite having a
misconceived notion of how parameters are passed to functions
\cite{madison02}.

At a more general level, novice students may not internalize correctly
what they have learned.
As a result they may misapply it.
For example, Jeffries et al.\ report on students who used large arrays
instead of linked lists to store page numbers of an index, and one who
decided to store the text to be indexed in a binary tree, thus losing
all its structural information, instead of storing the desired index
terms in a tree \cite{jeffries81}.

The conclusion from studies such as these can be summarized as
follows:
\mybox{Novice programmers,
  especially beginning undergraduate students, may be ill-equipped for
  participation in empirical software engineering studies in fields such
  as program comprehension or design.
  At the very least, more advanced students (e.g.\ from the third year
  of study or beyond) should be preferred.}

\subsection{Differences in Technology Use}

Another difference between students and more experienced developers is
how they use available technology and tools.
Several studies have documented such effects, and interestingly, they
work in both directions.

Inexperienced users may not use available technology that could help
them to perform a task, either because they don't know about the tools
or they do not feel sure enough to use them.
As an example, inexperienced students faced with the task of
comprehending Java code that had been obfuscated by identifier
renaming did not use renaming facilities to change the obfuscated
names back to meaningful names once they figured out what they
represented; instead they continued to work with the original
obfuscated code \cite{ceccato14}.
Thus they did not use an available tool that could help them.

On the other hand, students also stand to gain more from using tools,
because such tools can make up for the shortcomings in their
experience.
For example, Ricca et al.\ demonstrated that inexperienced students
benefited more from using web-application-related annotations on UML
diagrams than more experienced subjects \cite{ricca10}: the
experienced ones could understand the diagrams to a large degree even
without the added annotations, but for inexperienced students the
added annotations proved invaluable.
In other words, the tool --- in this case, the added annotations on
the UML diagrams --- helped the inexperienced users to close the gap
between them and the more experienced users.

But this can also work the other way around.
Tool usage takes training, and the learning curve itself may depend on
previous knowledge.
Thus advanced tools may be accessible only to advanced users, and by
using them the advanced users \emph{widen} the gap between them and the
inexperienced users.
An example is provided by Abrah\~ao et al.\ \cite{abrahao13}.
The context was again the use of UML diagrams, in this case sequence
diagrams used to portray and understand system dynamics.
Experiments showed that using these diagrams helped experienced and
proficient users comprehend the modeled functional requirements, but
had no significant effect for inexperienced users.
Conversely, students may actually be better trained in tool usage than
professionals who have not been formally taught the basics
\cite{arisholm06}.

Briand et al.\ found that adding formal OCL (Object
Constraint Language) annotations to UML diagrams helped students
perform comprehension and maintenance tasks, but only provided ample
training was provided first \cite{briand05}.
Such training was also needed for the effective use of a graphical
representation of requirements \cite{sharafi13}.
Another example was described by Basili et al.\ \cite{basili99}.
In reviewing five replications of experiments on the effectiveness of
defect-based code reading (DBR) as a means to uncover faults, three
studies based on undergraduates showed no evidence for increased
effectiveness, while two based on graduate students and practitioners
showed evidence for a positive effect.
This was attributed to the fact that DBR is based on expressing
requirements in a state-machine notation, which may be too
sophisticated for undergraduates.

An especially interesting experiment shows that differences in tool
usage may represent differences in strategy.
Bednarik and Tukiainen used eye tracking to see how study participants
performed program comprehension tasks, based on code reading and
animations of the program execution using the Jeliot 3 visualization
tool \cite{bednarik06}.
The result was that some participants first read the code to form a
hypothesis of what it does, and then used one or a few animations to
verify this.
These participants turned out to be the more experienced ones, with
$73.5\pm69.9$ months of programming experience.
Other participants, on the other hand, approached the comprehension
task by running animations to see what the programs do.
It turned out that these participants could easily be characterized as
less experienced, having only $18.0\pm19.4$ months of experience.

The conclusion from these examples is 
\mybox{Study subjects should be selected to fit the task.
  Studies involving sophisticated tasks, which require mastery of
  specific tools or methodologies, should not employ student subjects
  who do not have the necessary skills.
  At the very least, appropriate training should be provided, and the
  level of proficiency should be assessed.}

We note in passing that experience may not only affect tool usage, but
may also inform tool design.
In a series of studies LaToza and Myers have surveyed professional
developers, and recorded the most vexing problems that troubled them
during their work \cite{latoza10b,latoza10a}.
They then used this to design tools that specifically target these
problems.
Importantly, the uncovered issues were hard to anticipate, as they
reflected problems and work practices of professionals in a real-life
setting.
For example, the developers often faced problems such as understanding
why a certain piece of code was implemented in a certain way, what
code does under certain conditions, why it was changed, and so on.
Students typically do not face such questions, and can not be
expected to identify them as important.

\subsection{Lack of Experience}

Students who are new to the field by definition lack experience,
meaning that they have not done much yet.
The interactions of experience with tool usage cited above are not the
only effect of experience.
Experience has been shown to have an effect on myriad other software
engineering activities as well.

At a rather basic level, experience may affect how one performs even a
simple task.
For example, Burkhardt et al.\ used a documentation task to track the
mental models of expert and novice programmers \cite{burkhardt02}.
One of the findings was that experts put three times as many comments
in header (.h) files as in code (.cc) files; for novices it was
exactly the other way around.
Moreover, within code files experts emphasized comments on functions
as opposed to inline comments (51\% vs.\ 29\%, respectively), whereas
for novices the inline comments outnumbered the function comments by a
2:1 ratio.
It can be conjectured that such behavioral differences stem from
experience in trying to understand others' code, and the realization
that general explanations and interfaces are more important than code
minutiae.
In a related vein, several studies have found that students place more
emphasis on comments than professionals do \cite{mantyla09,salviulo14}.

Several additional recent experiments have demonstrated situations in
which experience has a considerable effect on outcome.
Binkley et al.\ conducted a few studies of the effect of multi-word
identifier styles (that is, using camelCase or under\_scores) on their
correct and speedy recognition.
Initially they found that in general it took longer to identify
camelCase and it was faster to identify under\_scores
\cite{binkley09b}.
However, computer science training led to a reduction in the
difference: more training led to reduced time with camelCase and more
time for under\_scores.
All the subjects were undergraduate students, and the level of
training reflected years of studying computer science;
importantly, they also included non-computer-science majors as a
control group.

These results were largely replicated by
Sharif and Maletic \cite{sharif10}, 
using eye-tracking equipment and subjects trained mainly in the
under\_score style.
The subjects were both undergraduate and graduate students and two
faculty members.
In another experiment, this time measuring the ability to recall
identifiers, beginners did badly when the identifiers had
under\_scores \cite{binkley13}.
Thus the experts did better than beginners with under\_scores, and
beginners did better with camelCase than with under\_scores.
But these studies also showed many other effects, that in some
cases were much more significant than the effect of experience.

Considering more complicated tasks, a study of 5 professional
developers by Adelson and Soloway showed that experienced developers
may use different approaches depending on whether they have previous
domain experience that is specifically relevant to the problem at hand
\cite{adelson85}.
Novices don't have such resources available to them, and need to build
up their knowledge of the domain from scratch.
But gaining experience may be quick.
Falkner et al.\ document how students change their software
development strategies from their first year to the end of their
degree \cite{falkner15}.

It is not surprising that experienced developers perform better than
novice programmers on various programming-related tasks.
But what is the root cause of this advantage?
Soloway and Ehrlich suggested that expert advantage stems from two
reasons: they know solution patterns, and they follow and exploit
programming conventions \cite{soloway84}.
Experiments to demonstrate this were conducted with 145 students, with
a 2:1 ratio of novices to advanced ones (in their case, students with
at least 3 programming courses under their belt), using tasks that
were meant to specifically depend on these two traits.
The results were that indeed the advanced students performed better.
But when program fragments did not conform to the patterns and
conventions, the performance of the advanced programmers was reduced
to near that of novices.
More results on the internal representations used by novices and
experts are described below in Section \ref{sect:dir}.

In summary, the above results support and extend those of the previous
section:
\mybox{Experience may certainly contribute to capability and
  change how problems are tackled.
  Therefore experimental subjects should have a suitable level of
  experience to undertake the tasks being asked of them.}

\subsection{Academic Overqualification}

As mentioned already one of the justifications for using students in
empirical software engineering research is the perception that they
are representative of the next generation of software engineers
\cite{kitchenham02,tichy00}.
his notion is challenged by the Stackoverflow 2015 Developer Survey%
\footnote{http://stackoverflow.com/research/developer-survey-2015,
  visited 29 Jul 2015.}.
Slightly more than 20,000 developers worldwide answered the survey.
Among them, 37.7\% had a BSc in computer science or a related field,
18.4\% had an MSc, and only 2.2\% had a PhD.
These numbers are rather similar to those reported in a survey of
developers using Microsoft technology reported in VisualStudio
Magazine in 2013%
\footnote{https://visualstudiomagazine.com/articles/salary-surveys/salary-survey.aspx,
  visited 29 Jul 2015.};
The numbers there were 33.9\% who graduated from a 4-year college,
28.0\% with an MSc, and 3.0\% with a PhD.
However, an additional 10.6\% graduated from a 2-year college, and an
additional 8.1\% took some post-graduate studies without obtaining a
degree.

Returning to the Stackoverflow survey, 48\% reported that they did not
have a university degree in computer science, and 33\% reported that
they did not attend even one university computer science course.
In fact, 41.8\% claimed to be self taught.
Thus computer science students (and the university education they
receive) represent at best about half of developers in the field, and
graduate students are especially non-representative.
In other words, the notion that students are at least representative
of the next generation of software engineers (if not of the current
generation) may be optimistic.

A different view on this issue is presented by Lethbridge, who
conducted a survey of computer science and software engineering
graduates in 1998 \cite{lethbridge98}.
He found that respondents generally considered their education to be
moderately relevant to their work as software developers, but with a
wide range of opinions, including some who thought it was completely
irrelevant.
More specifically, of those who studied computer science or software
engineering 70\% considered their education relevant, but of those who
had studied computer or electrical engineering only 30\% found it
relevant.
More interestingly, respondents reported having learned significantly
more mathematics than software, but software outranked mathematics by
about the same margin for usefulness, current knowledge, and desire to
learn more.
This implies that academic studies are not perfectly aligned with the
needs of industry professionals.
Thus,
\mybox{A little-appreciated problem with students is that their
  perspective is academically oriented, and perhaps not aligned with
  industry practice.}

\subsection{Contextual Effects}

The previous sections dealt with technical aspects of how students
may perform in software engineering experiments, and how this may
differ from the performance of professionals.
But there are also concerns that stem from the fact that students are
just students and that the experiments are taking place in an academic
setting.

One possible problem is uniformity and lack of perspective.
In an academic setting, it is highly probable that all the student
subjects had learned the relevant material from the same professor at
the same time.
Furthermore, they most probably did not accrue any on-job experience
that could challenge what they had learned and confront it with
day-to-day situations.
Thus their collective perspective on the issue being studied may be
limited.

Concern has also been expressed about the effect of the
relationship between students and (faculty) experimenters
\cite{ko15}.
This may take either of two forms: the students may anticipate the
desired outcome and unconsciously favor it, or they may sense their
professor's intent and unconsciously try to support it.
While such effects have apparently not been reported in the context of
software engineering, they are known from other domains.
Another potential danger occurs when students are required to
participate in an experiment and are not allowed to withdraw at will.
In such a situation there is a risk of contaminating results by
disgruntled subjects \cite{singer02}.
But this is not necessarily limited to students, and may also happen
when professionals are drafted to participate in a study by their boss.

Another concern is that students may be less committed than
professionals \cite{host00,berander04}.
This may be especially significant if the experiments are done as part
of a class, as in this setting students (at least the smart ones) may
be expected to invest effort only to the degree that it contributes to
their grades.
Volunteer students or those participating in a project may therefore
make better experimental subjects.
But on the other hand, it is not clear that professional participants
are necessarily more committed, especially if they are not paid
specifically to participate in the experiment.
For example, one study describes a situation where over 100 employees
in a firm agreed to participate in a study, only 60 eventually
submitted their results, and of those only 33 were really
usable \cite{causevic13}.
Another study also noted lack of cooperation as a problem, and
specifically cited the inclination of industry participants not to follow
experimental guidelines, sometimes voiding the results simply because
they did not actually use the prescribed methodology that was being
examined \cite{vegas15}.
\mybox{Conducting experiments in an academic setting may be suspect.
  However, there are no specific reports of such problems in software
  engineering research, and it is not clear that an industrial setting
  is better.}

\section{Direct Comparisons of Students and Professionals}
\label{sect:dir}

Many of the experiments described in the previous section considered
experience as a confounding factor and not as a main effect.
But because of concerns about using students as subjects, some papers
have also addressed the issue of comparing students to professionals
explicitly.
Such papers are listed in Table \ref{tab:studprof}, and those that
have not been mentioned yet are discussed further next.
In this we expand the discussion to include not only studies pitting
students against professionals, but more generally, any study
concerned with programmers with different levels of expertise.
This can include comparing the performance of novice students with
more advanced ones
(e.g.\ \cite{ceccato14,ricca10,abrahao13}),
or novice professionals with more experienced ones.
It can also involve a sort of meta-study, where a study with one group
of subjects is later followed by another study using another group,
and the results are compared (e.g.\ \cite{briand01b,causevic13,ceccato14}).

\subsection{Studies of Cognitive Processes}

In the 1980s there was some interest in the field of psychology in
understanding the differences between expert and novice programmers.
These studies were focused on cognitive processes and internal
representations.
The common theme was that experts employ deeper knowledge and
semantics, while novices tend to use surface attributes and syntax
\cite{shneiderman79}.
For example, Weiser and Shertz asked programming novices and experts
(undergraduates and graduate students) to classify 27 problems.
The novices tended to do so based on application area, whereas the
experts tended to classify based on the algorithm that would be used
to solve the problem \cite{weiser83}.

McKeithen et al.\ demonstrated that expert programmers are better able
to recall semantically meaningful program code, and that their
programming knowledge is better organized to reflect programming
concepts \cite{mckeithen81}.
To quantify this they asked subjects to recall 21 ALGOL keywords, each
time prompting them to start with a different one and continue with
others ``that go with it''.
They then analyzed subsequences that tended to appear in the same
order.
Wiedenbeck showed that experts are both faster and more accurate on
simple programming related tasks (e.g.\ to identify syntactically
incorrect lines), implying that experience leads to automation that
replaces conscious cognitive effort \cite{wiedenbeck85}.
These results hark back to the work of Simon on chess masters, who
were shown to possess a large ``vocabulary'' of chess positions
\cite{simon73}.

Adelson compared novice programmers with more advanced ones (graduates
of an introductory programming course and TAs in that course), using
materials and questions at two different cognitive levels
\cite{adelson84}.
The abstract level represented a program using a flow-chart with
high-level blocks, and the corresponding questions were also at this
level (e.g.\ what is the shape of a matrix being used).
The concrete level represented the program using a detailed flow-chart
of individual operations, and the questions were about details
(e.g.\ which border of the matrix is handled first).
The results were that novices tend to operate at the concrete level
and experts at the abstract level --- so much so, that in certain
cases presenting a concrete question regarding an
abstractly-represented program caused experts difficulties, and
allowed novices to out-perform them.

In a related vein, Burkhardt et al.\ studied the internal
representations of novices and experts for object-oriented programs
\cite{burkhardt97}.
The experts were 30 professional programmers, and the novices were 21
advanced students who were relatively new to object technology.
Using documentation and code reuse tasks, they found that the experts
had a better mental model of the objects that constitute the program
and the relationships between them.
However, novices could improve their representation if the task
demanded it \cite{burkhardt02}.

\begin{longtable}{>{\raggedright\arraybackslash}p{17mm}p{10mm}p{10mm}p{8mm}>{\raggedright\arraybackslash}p{45mm}>{\raggedright\arraybackslash}p{45mm}}
\caption[]{\label{tab:studprof}\sl
Papers comparing experimental subjects with different levels of
education or experience.
L1, L2, and L3 are the levels used in the experiments.
Numbers in parentheses are the number of subjects at each level.
``stud.'' is students of unspecified degree, and $\uparrow$ indicates
students near the end of their degree.
``ind.'' or ``prof.'' means professionals from industry.
If degrees and affiliation are not mentioned assignment was made by
some assessment of proficiency and classification as ``beginner'',
``novice'', ``intermediate'', ``advanced'', ``experienced'', etc.} \\
\hline
& \multicolumn{3}{c}{Subjects} &&\\
\cline{2-4}
\emph{Ref.}&\emph{L1}&\emph{L2}&\emph{L3}&\emph{Study}&\emph{Results} \\
\hline
\endfirsthead
\caption[]{Continued.} \\
\hline
& \multicolumn{3}{c}{Subjects} &&\\
\cline{2-4}
\emph{Ref.}&\emph{L1}&\emph{L2}&\emph{L3}&\emph{Study}&\emph{Results} \\
\hline
\endhead
\multicolumn{6}{r}{(Continued on next page)} \\ \hline
\endfoot
\hline
\endlastfoot

Sackman \cite{sackman68} 1968
& trainee (9) & exp. (12) & -- &
verify the effect of an interactive environment on debugging &
for both groups the interactive setting was beneficial, and for both
individual differences were high \\
\hline

Jeffries \cite{jeffries81} 1981
& nov. BSc (5) & exp. (4) & -- &
compare solution strategies to book indexing problem &
all subjects used the same decomposition strategy, but novices were
less effective in looking for subproblem solutions and knowledge
representation \\
\hline

McKeithen \cite{mckeithen81} 1981
& nov. stud. (29) & int. stud. (29) & exp. (8) &
identify how programming knowledge is organized depending on skill &
more advanced subjects' knowledge was better organized to reflect
programming concepts \\
\hline

Weiser \cite{weiser83} 1983 
& nov. (6) & exp. (9) & mgr. (4) &
classify problems by domains, algorithms, or data structures &
novices tended to classify by domain, experts by algorithm, and managers were
different from both \\
\hline

Adelson \cite{adelson84} 1984 
& nov. stud. (42) & TAs (42) & -- &
identifying the cognitive level at which novices and experts work &
novices use concrete representations and experts use abstract
representations \\
\hline

Soloway \cite{soloway84} 1984 
& nov. stud. (94) & adv. stud. (45) & ind. (41) &
characterizing the knowledge of advanced programmers &
advanced programmers know and use design patterns and programming conventions \\
\hline

Wiedenbeck \cite{wiedenbeck85} 1985 
& nov. (20) & exp. (20) & -- &
test performance on trivial programming-related tasks &
experts are faster and more accurate, implying automation \\
\hline

Gugerty \cite{gugerty86} 1986 
& nov. (10) & exp. (10) & -- &
finding a bug in a program &
skilled programmers are faster and better \\
\hline

Basili \cite{basili87} 1987 
& stud. (42) & ind. (32) & -- &
compare software testing and inspections &
code reading worked better for professionals \\
\hline

Porter \cite{porter95,porter98} 1998 
& grad. (48) & ind. (18) & -- &
systematic strategies for requirements inspections &
comparisons of approaches gave the same results despite differences in
absolute performance \\
\hline

Schenk \cite{schenk98} 1998 
& nov. (7) & exp. low (9) & exp. high (9) &
detailed study of how system analysts do requirements analysis &
experts approach requirements analysis differently from novices \\
\hline

H\"ost \cite{host00} 2000 
& MSc$\uparrow$ (25) & ind. (17) & -- & 
subjective assessment of project delay factors &
similar results for both groups \\
\hline

Briand \cite{briand01b} 2001 
& stud. & ind. & -- &
use defect data to evaluate OO quality metrics &
results from student projects largely confirmed by industrial followup \\
\hline

Crosby \cite{crosby02} 2002
& nov. (9) & adv. (10) & -- &
how experts and novices work to comprehend code &
experts focus more on the complex parts \\
\hline

Burkhardt \cite{burkhardt97,burkhardt02} 2002 
& stud. (21) & ind. (30) & -- &
mental models of object-oriented programs &
complex interactions between factors, including some effect of expertise \\
\hline

Runeson \cite{runeson03} 2003 
& BSc$\downarrow$ (31) & grad. (131) & -- &
improvement due to PSP training &
freshmen and graduate students showed similar improvement, but
graduates were faster \\
\hline

Berander \cite{berander04} 2004 
& MSc$\uparrow$ (20) PhD (15) & proj. (16) ind. & -- &
requirements prioritization &
classroom students divide requirements more evenly between priorities;
those in industrial projects make most requirements high priority \\
\hline

Arisholm \cite{arisholm04} 2004 
& stud. jun. (90) & int. sen. (68) & -- &
maintenance of either a (bad) centralized or (better) delegated OO design &
inexperienced subjects had a hard time with the delegated style,
students did better on the centralized design \\
\hline

Bednarik \cite{bednarik05} 2005 
& nov. (8) & int. (8) & -- &
gaze analysis when using animations for comprehension &
advanced participants exhibited same behavior but were faster \\
\hline

Lange \cite{lange06} 2006 
& stud. (111) & ind. (48) & -- &
effect of defects in UML diagrams on their use &
similar effect on both students and professionals \\
\hline

Arisholm \cite{arisholm07} 2007 
& jun. (81) & int. (102) & sen. (112) &
compare individual programming with pair programming &
pairs of novices got better results on a complex system; experienced
pairs were slightly faster on a simple system \\
\hline

Bishop \cite{bishop08} 2008 
&stud. (34) & ind. (13) &  -- &
debugging spreadsheets &
experts were more methodical and found more bugs \\
\hline

Bannerman \cite{bannerman11} 2002--2008 
& stud. & ind. & -- &
metastudy about test-first vs.\ test-last development &
with students test-first was typically faster but with no quality
effect, with professionals it was slower and improved quality \\
\hline

McMeekin \cite{mcmeekin09} 2009 
& BSc$\uparrow$ (36) & ind. (26) & -- &
defect finding using 3 inspection techniques &
same relative performance of techniques;
different speed depending on experience \\
\hline

M\"antyl\"a \cite{mantyla09} 2009 
& stud. (87) & ind. & -- &
defect types found in code reviews &
similar high-level distribution of defect types were found \\
\hline

Ricca \cite{ricca10} 2010 
& BSc & grad. & RA &
effect of using web application annotations on UML diagrams &
added annotations were helpful for inexperienced students \\
\hline

Bergersen \cite{bergersen12} 2012 
& stud.\footnote{Cited previous work.} (266) & ind. (65) & -- &
debugging iterative vs.\ recursive code &
low-skill subjects did better on the recursive version, skilled ones
did the same on both \\
\hline

Soh \cite{soh12} 2012 
& stud. (12) nov. (9) & ind. (9) exp. (12) & -- &
understanding and maintaining UML diagrams &
practitioners were more accurate, but for a given accuracy students
were faster \\
\hline

\v{C}au\v{s}evi\'c \cite{causevic13} 2013 
& MSc (14) & ind. (33) & -- &
test cases in test-driven development &
similar behavior in both groups \\
\hline

Abrah\~ao \cite{abrahao13} 2013 
& low & high & -- &
effect of UML sequence diagrams on comprehension &
sequence diagrams helped high-ability participants \\
\hline

Jung \cite{jung13} 2013 
& PhD (7) & ind. (11) & -- &
compare 2 safety analysis tools &
replication in industry gave essentially same result of no difference
between tools \\
\hline

Ceccato \cite{ceccato14} 2014 
& BSc (13) & MSc (39) & PhD (22) &
the effect of code obfuscation on comprehension &
more advanced programmers are better but affected in a similar way \\
\hline

Salviulo \cite{salviulo14} 2014
& BSc$\uparrow$ (18) & prof.$\downarrow$ (12) & -- &
the use of comments and variable names for comprehension and maintenance &
students used comments more, all agreed that good identifier names are
important \\
\hline

Daun \cite{daun15} 2015 
& BSc (125) & MSc (21) & -- &
review specifications of an avionics system in either of two ways &
graduate students were more effective leading to larger effect
size, but results for undergrads were also significant \\
\hline

Salman \cite{salman15} 2015 
& grad. (17) & ind. (24) & -- &
effect of using TDD on code quality &
both groups produced code of similar quality when using TDD for the
first time \\
\hline

Busjahn \cite{busjahn15} 2015
& nov. (14) & ind. (6) & -- &
use eye tracking to study the order of code reading &
code reading is less linear than story reading, and experts are less
linear than novices \\

\end{longtable}

At an even higher level, Jeffries et al.\ compared five undergraduate
students to four experts (an EE professor, two experienced graduate
students, and a professional) on how they actually designed a
program to create an index for a book.
All subjects used the same general strategy of decomposing the problem
into subproblems.
However, the experts were much more effective.
Thus the novices tended to make do with the first solution they found
for a problem and did not consider alternatives, they missed or
decided to ignore subtle end cases, and they did not rely on using
known algorithmic solutions.

Putting all of this together,
\mybox{It appears that expert programmers think differently from
  novices.
  This may hinder the use of students, especially at the start of
  their studies, for tasks that are conceptual or presented at a high
  level of abstraction.}

\subsection{Studies Showing Similarity of Students and Professionals}

Quite a few studies have compared the performance of students and
professionals on concrete programming-related tasks.
An especially interesting experiment was conducted by
McMeekin et al.\ \cite{mcmeekin09}.
The goal was to compare three software inspection techniques:
checklist-based reading (CBR), which was developed in the context of
procedural programming, and use-case reading (UCR) and usage-based
reading (UBR), which were developed for object-oriented inspection.
The reported results were that there was no significant difference
between the inspection techniques, but there was indeed a significant
difference between the 36 students and 26 professional developers
applying them.

However, the paper also included a relatively detailed rendition of
the raw results, in the form of box plots of the distributions of
number of defects found by students or industry professionals using
each of the 3 methods.
Except for the scale (professionals found about twice as many defects)
these two graphs are extraordinarily similar, including details such
as the following:
\begin{itemize}\itemsep 0pt
\item The distribution for CBR is the least disperse, and for UBR the
  most disperse.
\item The distribution for UBR subsumes the distribution
  for CBR: the plotted percentiles above the median were higher for
  UBR, and those below the median were lower for UBR.
\item The distribution for UCR seems to be shifted slightly toward
  lower values relative to the other two.
\item For CBR the median is closer to the 25th percentile, and for UCR
  it is closer to the 75th percentile.
\end{itemize}
Thus a reinterpretation of the results is that the experiment with
students actually provides an excellent approximation of the results
obtained with professionals, as far as the comparison between the
three techniques is concerned.

Several other studies also found that the behavior of subjects with
different levels of experience was very similar in everything but
speed.
The earliest is the study by Sackman et al.\ from 1968 regarding
interactive debugging, which showed that both trainees and experienced
professionals benefited from using an interactive setting
\cite{sackman68}.
Moreover, both groups also exhibited similar individual differences
between the group members, which was somewhat surprising for the
trainees which were expected to be more similar to each other.
Bednarik et al.\ looked at the gaze behavior of novice
and intermediate programmers when using code and animations during
program comprehension \cite{bednarik05}.
The only difference was that the more advanced participants did
everything faster, except for the control functions (clicking on
buttons to control the animation) where the speed was the same.
This exception may actually lend credence to the results, by
suggesting that the difference between the subjects was limited to
programming experience, and did not extend to button-clicking dexterity.
As another example, Porter and Votta studied systematic strategies for
requirements inspections, first using graduate students and then with
industrial professionals \cite{porter95,porter98}.
Again the numbers were different, but the statistical tests comparing
different approaches told the same story.

M\"antyl\"a and Lassenius used both students in a course and
professionals in actual work to study the types of defects which are
found during code inspections \cite{mantyla09}.
For both groups, the vast majority (77\% and 71\% respectively) were
found to be evolvability defects, and only a minority (13\% and 21\%)
were functional; the remainder were false positives.
However, there were some differences in the details.
For example, professionals noted more defects concerned with structure
and naming, while students noted more defects in code comments.

Similar correspondence between students and professionals was found by
\v{C}au\v{s}evi\'c et al., who initially conducted a study on test-driven
development using 14 students.
This found that two thirds of the test cases that they produced were
positive, but around 60\% of the defects were found by the third of
the test cases that were negative.
This gap was then confirmed in an industrial followup with 33
developers, where only 29\% of test cases were negative but they were
responsible for finding 71\% of the defects \cite{causevic13}.
Briand et al.\ used a similar methodology, where results from student
projects were subsequently largely confirmed by an industrial followup
\cite{briand01b}.
The context was using defect data to evaluate object-oriented quality
metrics.
Another such replication was done by Jung et al., in which two
security analysis tools were shown to support essentially the same
level of performance \cite{jung13}.

In the same vein,
Svahnberg et al.\ report on a study conducted in the
context of a requirements engineering course, where students were seen
to be able to anticipate the values of industrial practitioners
\cite{svahnberg08}.
Specifically, the students correctly thought that practitioners will
value the business perspective significantly more than the system
perspective, despite not making this distinction themselves.
Daun et al.\ also used students in requirements engineering courses
\cite{daun15}.
They found that while graduate students were more efficient (faster)
and better at identifying stakeholder intention, undergrads produced
the same qualitative results.
Lange and Chaudron found that students and professionals
were affected in a similar manner when they needed to work with
defective UML diagrams \cite{lange06}.

A rather different type of study was conducted by
H\"ost et al.\ \cite{host00}.
They used 25 students (last year software engineering MSc) and 17
professionals (with an average of 11 years experience).
But the task was not a conventional development or maintenance task,
but rather a subjective assessment of the relative importance of 10
factors which may affect the time needed to complete a project.
It turned out that the assessments of the students and the
professional were quite similar, leading to the conclusion that
students may be used in lieu of professionals in this case.
This is especially interesting because the task involved judgment and
not technical skill.

Similarity between students at different levels was also found in how
they benefited from novel training procedures.
Runeson checked the improvement of freshmen students and graduate
students who received PSP (Personal Software Process) training, and
also compared this with professionals who underwent similar training
\cite{runeson03}.
The results were that all groups exhibited similar improvements in 7
factors that were quantified.
The only exceptions were that freshmen improved much more than the
others in estimation accuracy and productivity, perhaps because they
started at a much lower level.
In a related vein, a study of the code quality produced by graduate
students and professionals when using a new technique, in this case
test-driven development, found that the quality was similar for both
groups, despite being different when using traditional methods for
which they had experience \cite{salman15}.

The similarity of professionals and students also has an ironic twist.
In a study about obfuscation, identifier renaming to obfuscate
meaningful names was hard for everyone (both experienced and
inexperienced subjects) in similar degrees.
This indicated that ``implicit documentation'' as reflected in
meaningful names has a stronger effect on comprehension than
experience \cite{ceccato14}.

To summarize,
\mybox{In many studies students were shown to produce the same
  relative results as professionals, allowing for correct ranking of
  alternative treatments.
  However, professionals were typically faster or more effective.}

As noted, in practically all the studies cited here the practitioners
were found to be faster than the students, sometimes by a wide margin.
But the opposite is in principle also possible.
For example, Schenk et al.\ found that expert information systems
analysts took more time and verbalized way more considerations than
novices when performing a system analysis task \cite{schenk98}.
Likewise, Atman et al.\ found that professional engineers take more
time than students to solve a given problem, because they spend more
time on problem scoping and on information gathering \cite{atman07}
(albeit this was in the context of designing a playground, not
software).
Such results are representative of studies which found differences
between students and professionals.

\subsection{Studies Showing Meaningful Differences}

We already saw some examples of studies that demonstrated meaningful
differences between students and professionals in Section
\ref{sect:stud}.
Here we review a few more which identify specific elements of
expertise which may be missing in students.

A relatively early study was conducted by Basili and Selby to compare
the relative effectiveness of code inspections and testing schemes
\cite{basili87}.
Their results show that code reading leads to better results for
professionals, but not for students.
An especially noteworthy aspect of this study is that
they also classified their subjects according to their level of
expertise based on academic performance and years of experience; for
professional subjects their manager's opinion was also taken into
account.
Among the students, 29 were thus classified as ``junior'' and 13 as
``intermediate''.
Among the professionals, 13 were ``junior'', 11 were ``intermediate'',
and 8 were ``advanced''.
But the results were that performance differences between
professionals and students (irrespective of experience) were more
consistent than performance differences between subjects with
different expertise levels.
This is somewhat tempered by the fact that differences between
expertise levels were rather noisy, possibly because the assignment
into expertise levels was not based on a direct evaluation.

Comparable results were obtained by Soh et al., in a study where
the understanding and maintenance of UML diagrams was used as the
experimental task \cite{soh12}.
This study involved 12 students and 9 practitioners.
But in addition to comparing them based on status, they were also
compared based on their years of experience, and the result in this
case was that years of experience was the more important factor.
However, this result is somewhat tainted by the relatively small
number of subjects and by the large overlap between the
classifications: 8 of the 9 practitioners were considered experienced,
and 8 of the 12 students were novices.

Berander reports on student behavior that deviates from the observed
behavior in industry in the context of requirements prioritization
\cite{berander04}.
This was done by conducting a prioritization game, where clients set
priorities, developers provide time estimates, and together they draw
a schedule of how the planned features will be distributed across
product releases.
Observations from industry, and also from students working on a
large-scale project with industry partners, shows that there is a
clear tendency to pile the vast majority of the requirements into the
``most important'' category.
But when students played the role of clients in a classroom study,
they overwhelmingly tended to divide the requirements evenly between
the different priorities.
This simplifies the scheduling but is unrealistic.

A possible generalization of behavior is that experts are more
methodical \cite{jeffries81}.
In a non-software example, professional engineers took more time than
students to design a playground, and the extra time was spent mainly
on problem scoping and information gathering \cite{atman07}.
In debugging a spreadsheet, professionals covered more of the cells
more consistently \cite{bishop08}.
But the professional outlook can also be a disadvantage: in the
debugging tasks, professionals were found to be more tuned to logic
and concepts, so they tended to miss superficial issues like a typo in
a constant.

Lui and Chan give a striking example where the level of expertise may
confound the experimental results \cite{lui06}.
Their study was concerned with the effectiveness of pair programming.
Many studies have been conducted on this topic, with mixed results.
The novelty of Lui and Chan's work was in controlling for the level of
the participants.
They conclude that pairs of novices, or in general pairs of
programmers confronted with new and challenging problems, stand to
benefit considerably from pair programming.
But for pairs of experts, or generally pairs confronted with problems
they have handled before, this is a waste of effort.
The experiment showing this was based on repeat-programming: the
subjects were asked to solve the same problem time after time, and the
benefit of working in pairs was seen to drop with iteration number.

Arisholm and Sj{\o}berg designed an experiment showing how mixing
people with different levels of competence can lead to problems in the
field \cite{arisholm04}.
Specifically, they used two different design styles to solve the
problem of implementing a simple coffee machine: one in which the
front panel object is in control and just uses other object to perform
simple tasks, and the other in which the front panel merely initiates
activities and delegates the actual execution to other objects which
encapsulate the details.
The delegated style was considered the better object-oriented design.
In the experiment, undergraduate students and junior consultants were
found to have problems maintaining the delegated style, whereas
graduate students and senior consultants performed better with this
style.
Thus if senior designers are employed and use the preferred style to
design a product, there is a danger that more junior maintainers will
be ill-equipped to maintain it.

In followup work with additional colleagues, Arisholm et al.\ compared
the above results (but excluding the students) with the work of pairs
of professional developers \cite{arisholm07}.
The results were that working in pairs was in general not beneficial,
as the strongest effect was an 84\% increase in total effort (because
two people were involved and it took approximately the same time).
However, they did observe an interaction with problem complexity:
pairs of junior and intermediate consultants achieved significantly
better results than individuals on the complex (delegated) problem,
and pairs of intermediate or senior consultants took somewhat less
time than individuals on the simple (centralized) version.

It is hard to generalize the above results.
But we can state that
\mybox{Many factors affect the performance of software engineering
  tasks, and experience or expertise sometimes figure among them.
  However, this often interacts with other factors, and pinpointing
  the precise effect is difficult.
  Many more experiments on diverse aspects of this issue are desirable.}

\section{Discussion}
\label{sect:disc}

As shown in the previous sections, there is a wide range of studies
which have considered --- or grappled with --- the confounding effects
of experience on software engineering performance.
But can this be limited to a discussion of using student subjects as
opposed to professionals?
And what other considerations should be applied?
In the following we discuss these and related questions based on the
evidence reviewed above and additional literature from other fields
concerning topics like experience and expertise.

\subsection{The Semantics of ``Student'' and ``Professional''}

The basic complaint against using students as experimental subjects is
that they are perceived as not representative of professional
developers in a ``real'' industrial setting.
But real life may be more complicated, leading to situations where
this distinction does not correspond with the intent.

First, it should be noted that ``students'' come in many varieties.
Undergraduate students are not the same as graduate students.
Within the set of undergraduates, freshmen (first year students) are
different from second or third year students.
Likewise, within the set of graduate students, a distinction can be
made between masters students and doctoral students.
In fact, many studies compare undergraduate students to advanced
graduate students when they want to assess the effect of experience,
or else use students from different levels to control for the effect
of experience \cite{abrahao13}.
In addition, there may be differences between students studying in
different degree programs (e.g.\ software engineering vs.\ computer
science) and at different institutions (e.g.\ a community college
vs.\ a technical university).

Second, the dichotomy assumes that students lack professional
experience.
This is not always the case.
Many youngsters nowadays learn to program in their teens, and can
accumulate quite a lot of experience in hacking or contributions to
open source projects before they graduate from high school.
Some may even establish their own startup companies.
In Israel, for example, many students come with experience from their
army service \cite{senor:book}.
In some study programs internships in industry are incorporated into
the program, thereby giving the students some professional experience
as they learn.
Students may also work in parallel to their studies in order to
support themselves;
thus study participants identified as coming from industry may
actually also be students, and even undergraduates
(e.g.\ \cite{maia13}).
In all these cases, the students can therefore actually have
significant industrial experience.

And third, the term ``professional'' is also not well-defined.
Different terms are sometimes used, such as ``developers'',
``practitioners'', or ``engineers'', and it is not clear exactly what
they mean and whether they are indeed different \cite{dieste13}.
Their training is also not necessarily the same;
For example, Arisholm and Sj{\o}berg report a case where senior
consultants in general held higher degrees than junior and
intermediate consultants \cite{arisholm04}.
Moreover, a professional can be new on the job or have the benefit of
many years of experience, just like students \cite{begel11}.
And finally, their jobs may actually be quite different too.
For example, are professional developers and testers the same?
Are those assigned to develop new functionality the same as those
assigned to perform maintenance tasks \cite{briand01}?
And what about having experience in different methodologies and
approaches?
While it stands to reason that experience in a specific relevant
domain may be beneficial to experimental subjects
(e.g.\ \cite{adelson85}), one can also argue for more general
capabilities that come with experience \cite{soloway84,sonnentag06}.

\mybox{The terms ``student'' and ``professional'' are ill-defined, and
  not necessarily exclusive.
  Using them to label experimental subjects may be misleading.}

\subsection{Novices and Experts}

Beyond the labeling of students and professionals, the more meaningful
objection is that students are \emph{novices}, and therefore do not
represent the work practices of professionals who are \emph{experts}.
So it is worth our time to consider these terms as well.

Experience is acquired over time by receiving feedback in real-world
situations.
Expertise is the result of internalizing this experience to create an
understanding of concepts and decision-making procedures, together
with an appreciation for potential biases and problems
\cite{schenk98}
(as also reflected in the Confucian saying ``I do and I understand'').
Novices by definition lack such expertise, and are therefore prone to
making errors in complex decision-making situations.

\begin{table}\centering
  \caption{\label{tab:novice-expert}\sl
  Characteristics of novices and experts, based on \cite{lord90,hmelos04}.}
  \begin{tabular}{l>{\raggedright\arraybackslash}p{55mm}>{\raggedright\arraybackslash}p{55mm}}
    \hline
    \emph{Feature}& \emph{Novice}	& \emph{Expert} \\
    \hline
    Search	& simplified heuristics until satisfactory alternative
		  found		& automatic homing in on best alternatives \\[2mm]
    Selection	& based on superficial features	& based on meaning \\[2mm]
    Information	& modest amount used	& extensive and highly organized \\[2mm]
    Focus	& system structure	& behavior and function of system elements \\[2mm]
    Process	& typically serial	& often parallel \\
    \hline
  \end{tabular}
\end{table}

The distinction between novices and experts is very important in any
field, and also in software development.
Experts know more and do things better, using different approaches
than novices.
Specifically, experts rely on deep domain knowledge to get to the
crux of the problem.
Novices are typically limited to relying on superficial cues.
In addition, it has been observed that novices treat complex systems
mainly at the structural level, whereas experts understand the
behavior and function of the system elements
\cite{batra92,hmelos04,hmelos07}.
This distinction may be expected to be particularly relevant to
software too.
Select attributes of the differences between novices and experts, as
reviewed by Lord and Maher, are listed in Table
\ref{tab:novice-expert}.

In the field of technical education more levels are identified.
An influential report on this issue was written by
Dreyfus and Dreyfus \cite{dreyfus80}.
The main levels they identified, and the characteristics of each
level, are
\begin{enumerate}\itemsep 0pt
\item Novice: knows to apply learned rules to basic situations
\item Competent: recognizes and uses recurring patterns based on experience
\item Proficient: prioritizes based on holistic view of the situation
\item Expert: experienced enough to do the above intuitively and automatically
\end{enumerate}
Thus the effect of experience is to enable the practitioner to break
away from prescribed rules and apply a wider perspective, eventually
doing this without conscious effort.
A fifth level, that of a master, has been suggested to reflect the
ability to innovate and develop new tools and methods to cope with new
situations.

While the difference between novices and experts may be quite
significant, the question of whether it is important in the context
of software engineering experiments is debatable.
One view is that expertise is not really important, for any of the
following reasons:
\begin{itemize}\itemsep 0pt
\item In many cases the tasks being performed by experimental subjects
  are not challenging enough, and do not require a holistic view or
  innovative solutions.
  Thus the differences between novices and experts come into play only
  during extended work on large-scale and difficult problems, and not
  in experiments.
\item Alternatively, when experiments involve technology or methods
  that are new to everybody, then skill or previous experience does
  not necessarily impart an advantage.
\item Finally, even if advanced professionals may do everything
  faster, and maybe even differently, still the qualitative
  experimental evaluation could well be the same as with students or
  other inexperienced subjects.
\end{itemize}

The other view is that the differences come into play even in simple
situations, as experts have a significant advantage in grasping the
situation and understanding the interactions between the code and the
context.
For example, Gugerty and Olson compared novices to skilled programmers
in a debugging task, and found that the skilled ones were faster,
found the bug more often, and hardly ever introduced new bugs while
hunting for the existing one \cite{gugerty86}.
This was attributed to the ease with which they were able to
dissect and encode the program, and thereby to comprehend the code and
come up with hypotheses about what was wrong.

Moreover, hardened professionals may acquire various practices
with time and experience that affect their approach, and cause them to
deviate from what may be expected and what students are taught.
For example, Roehm et al.\ conducted an observational study of how
professionals comprehend software when they need to perform a
maintenance task \cite{roehm12}.
A surprising result was that they may actually try to \emph{avoid}
comprehending the code, using various shortcuts instead.
For example, one participant identified which functions are relevant
by first commenting out all the applicable code, and returning the
functions one by one until there were no more compiler error messages.
As another example, Marasoiu et al.\ observed professional software
developers who used an environment's code completion feature as a
debugging aid --- if the code completion did not work, they took it as
an indication that something was wrong \cite{marasoiu15}.
So experts can leverage existing technology to their advantage in
unorthodox ways.

\mybox{There are real differences between novice and expert software
  developers.
  Being able to place experimental subjects on this spectrum is
  generally desirable, if only to see whether there exists an
  interaction between performance on the experimental task and
  expertise.}

\subsection{The Making of an Expert}

Using a title to describe experimental subjects (namely, labeling
them as ``students'' or ``professionals'') is often a stand-in for
asserting their level of expertise.
A possibly better alternative is to quote their \emph{experience}
(that is, years of doing software development),
and indeed many studies report that experienced developers perform
better (e.g.\ \cite{egan88,cheney84}).
For example, Robilard et al.\ studied how five developers approached
maintenance tasks on an open-source project (specifically, changes to
autosave functionality in jEdit) \cite{robillard04}.
They found that experience makes a big difference in success rates and
reflects real differences in approach.
Crosby et al.\ show that advanced programmers are somehow better able
to focus on the complex parts of the code \cite{crosby02},
and Busjahn et al.\ show that advanced programmers read code
differently from novices \cite{busjahn15}.
But in fact experience in years is also just an easy-to-measure proxy
for different levels of proficiency or capability.

Moreover, experience is not the whole story.
Another consideration is that there is evidence for very significant
variability among programmers.
Perhaps the first study to show this was the one by Sackman et
al.\ from 1968.
The goal of this study was to compare online and offline approaches to
debugging.
But one of its main findings was that the biggest differences in
performance were due to personal variability between experimental
subjects, which in one case reached a ratio of 28:1 \cite{sackman68}.
Large individual differences were also cited in various other studies,
e.g.\ \cite{myers78,curtis81,curtis89,klerer84,egan88,prechelt99,prechelt01}.
Personnel/team capability figures prominently on the cover of Boehm's
famous book \textit{Software Engineering Economics} as the cost driver
with the highest range of possible values, much more than any other
factor\footnote{The graphic on the cover (Figure 33-1 from the book)
  shows the range to be 1 to 4.18, more than double the range of 1 to
  2.36 for product complexity which is the next factor.
  But a factor of 4.18 is actually less than twice a factor of 2.36.}
\cite{boehm:econo}.
McConnell recounts an anecdote where a single programmer was called in
to replace a team of 80 (!) when a critical project was in risk of
missing a deadline \cite{mcconnell11}.
In the specific context of empirical research, such variability
could swamp out the effects being studied \cite{curtis14}.

Evidence for inherent differences in proficiency can also come from
computer science education.
In particular, a bimodal distribution of grades in CS 101 courses is
sometimes observed, and has led to discussions of the mechanisms that
create it.
One of the options is possession of a ``geek gene'', namely that some
people are predisposed to computer programming and therefore have an
edge on others \cite{ahadi13}.
More precisely, various concrete characteristics have been cited, such
as being able to clearly articulate a problem-solving strategy
\cite{simon06} and holding valid mental models of value and reference
assignments \cite{ma07}.
However, it is important to note that one of the more famous reports
suggesting that good programmers can be identified in advance based on
having consistent mental models of program behavior was later
retracted when more experimental data was collected \cite{bornat08}
and a replication also failed \cite{lung08}.
And other explanations have also been proposed, e.g.\ that a
fluke success early in the course can set a student on the better
learning path \cite{robins10}.

Based on the above, proficiency may be thought of as representing
the cumulative influence of three separate factors:
\begin{itemize}\itemsep 0pt
\item Natural talent.
  Personal differences between different people are most probably very
  important.
  Jackson has a wonderful short story equating ``brilliance'' with the
  talent to make things appear simple \cite[p.\ 20]{jackson:book}.
  Sex has also been shown to have some impact \cite{sharafi12}.
\item Formal training and deliberate practice.
  It may be expected (or at least hoped) that a university or college
  education improves one's proficiency, and that advanced degrees have
  some added benefit over that obtained from a first degree.
  The same goes for vocational training.
  And copious amounts of deliberate practice have been claimed
  to be the real factor leading to world-class performance
  \cite{ericsson93,ericsson07,colvin:talent}.
\item Experience on the job.
  This is easy to measure, at least superficially (that is, length of
  experience as opposed to quality of experience), but is at best only
  one of several factors that affect proficiency.
\end{itemize}
The conclusion is that
\mybox{Experimental studies should try to assess the
  actual proficiency of their subjects, and not just tag them as
  ``students'' or count their years of experience.}

\subsection{Experience and Proficiency}

Even if experience is only one factor contributing to proficiency, it
could still be that they are highly correlated.
It is not at all clear that experience is necessarily correlated with
proficiency in all cases:
developers with many years of experience may still be mediocre, while
relative novices may be very proficient.
Thus years of experience, while easy to measure, may turn out to be a
poor predictive attribute \cite{bergersen11}.
Also, even if more skilled developers may be more efficient in
term of time, their work may not necessarily be of higher quality
\cite{fucci15}.
\hide{
One reason for the above qualification is that other effects may be
more important than experience.
Talent and training are two such factors mentioned above.
Briand et al.\ studied the factors that influence programming effort,
by analyzing a dataset of space and military projects collected by the
European Space Agency \cite{briand99c}.
The result was that experience with the development environment
and programming language was a statistically significant factor in one
of the effort models.
However, its impact was small relative to the two main
factors: project size and team size.
}

It is also important to note that ``experience'' is by definition limited
to those domains and methodologies in which an individual has worked.
Thus it was found that professionals did not produce higher-quality
code than students when the experiment involved using a methodology
with which they did not have any prior experience, in this case
test-driven development \cite{salman15}.
Likewise, advanced students did not have an edge over novices when
confronted with obfuscated code, which was beyond the scope of their
previous experiences \cite{ceccato14}.
But strange counter examples also exist: Carver et al.\ have found
that inspectors with non-computer-science degrees found more
requirements defects than those \emph{with} computer science degrees
\cite{carver08}.
Another interesting and complicated interaction was discovered by
Basili and Selby \cite{basili87}.
They found that for professional developers code reading was more
effective than either black-box testing or statement coverage.
With students, however, they did not find such clear results, and in
one experiment all three methods were indistinguishable.
This result is especially noteworthy given that the professional
developers' main prior experience had been with functional testing,
not with code reading.
In other words, the advantage delivered by code reading depended in
some \emph{indirect} way on the experience of the professional
developers.

Sonnentag et al.\ provide a deep discussion of the differences between
\emph{experienced} developers and \emph{highly-performing} developers
\cite{sonnentag06}.
For example they claim that experienced developers tend to spend more
time than inexperienced ones on program comprehension and on
clarifying requirements, perhaps due to having experienced situations
where this was a problem (a similar trait has been observed for
engineers in general \cite{atman07}).
Highly performing developers, on the other hand, tend to spend
\emph{less} time on program comprehension, possibly because they
``get it'' faster.
An earlier study on high performers identified several differences
between them and more mediocre performers, but emphatically years of
experience was not one of them \cite{sonnentag98}.
However, it should be noted that these results pertain to the
comparison of professionals who were all experienced, and not to the
comparison of experienced professionals to novice students.

An interesting observation is that the effect of experience is not
linear.
Sackman et al.\ ran an early experiment of debugging time in online
vs.\ offline settings.
This included an initial assessment of aptitude before taking part in
the experiment itself.
The result was that the aptitude test had a good correlation with the
experiment results for trainees, but not so for professionals with an
average of 7 years experience.
In discussing this they speculated ``that general programming skill
may dominate early training and initial on-the-job experience, but
that such skill is progressively transformed and displaced by more
specialized skills with increasing experience'' \cite{sackman68}.
Thirty years later Sonnentag also showed that years of experience were
not correlated with differences in performance for professionals,
while citing results claiming that experience is indeed important for
students \cite{sonnentag98}.
This can be interpreted as suggesting that students are still
acquiring knowledge, whereas seasoned professionals have already
reached a plateau in their capabilities.
Similar observations were reported by Bergersen et al.\ \cite{bergersen11}.
These results agree with the ``laws of practice'', which state
that initially, when one lacks experience, every additional bit of
experience makes a significant contribution, but when one is already
experienced, the marginal benefit from additional experience is much
reduced \cite{newell81,heathcote00}.

\mybox{While experience may certainly contribute to capability and
  change how problems are tackled, it seems that using years of
  experience or employment are an overly simplistic metric for
  qualifications.
  In particular, beyond several years of experience individual
  differences probably become more important than differences in
  experience.}

At the very highest levels of performance, experience is not enough.
For example, the seminal work of Simon on chess masters indicated that
the most highly skilled needed to have invested a significant time in
obtaining experience, but not that anyone who invests such time
automatically becomes a master \cite{simon73}.
Similar findings have been uncovered in many other fields as well
\cite{ericsson93}.
Rather, deliberate practice is needed, meaning extensive repeated
practice directed specifically at those areas in which you are not yet
proficient, and directed by a mentor who provides candid feedback
\cite{ericsson07,colvin06,colvin:talent}.
While investing time is also required (ten years and 10,000 hours is
often cited as a minimum), it is how this time is used that is really
important.
But in an ironic twist of affairs, this also explains why such
top-notch expert performance is largely irrelevant in normal contexts:
deliberate practice is not fun \cite{ericsson93,colvin:talent}, so
most people do not engage in it.
In the specific context of software engineering experiments, we are
more often interested in studying ``normal'' practitioners than those
who have the exceptional capacity to push themselves to the limit.

\subsection{Assessing the Level of Proficiency}

Assessing the proficiency of experimental subjects is important for
several reasons \cite{kleinschmager11,myers78}:
\begin{itemize}\itemsep 0pt
\item First, it may enable us to claim that the results are relevant
  to a general setting, based on an evaluation showing that the
  subjects possess a representative level of proficiency.
\item Second, in the context of designing the experiment, we can
  verify that groups of subjects using different treatments are not
  biased in terms of ability.
  We may even use this to screen candidates so as to reduce the variance
  in capabilities, and avoid mixing subjects with different performance
  levels.
  Such mixing --- which occurs naturally due to the large variability in
  individual capabilities --- increases the dispersion of results for
  each treatment, and may mask the underlying differences due to the
  different treatments.
  This is especially problematic with small numbers of subjects.
\item Third, by quantifying the proficiency of different subjects, we
  can assess whether there exists an interaction between the
  experimental treatment and the level of proficiency
  \cite{bergersen12}.
  For example, this will tell us whether using a certain method or
  tool is more beneficial for proficient developers, for mediocre
  ones, or for novices.
\item Moreover, identifying the specific strengths and weaknesses of
  individual subjects \cite{acuna06} may uncover additional
  interactions and help avoid construct validity issues.
\end{itemize}

A Major issue is of course how to design tests with good
discrimination power, which will effectively identify and rank the
students (or other experimental subjects) on a scale from the most
proficient to the less so.
Moreover, software design tasks are typically ill-defined, meaning
that the problem specification is incomplete and discovered as part of
the solution process \cite{sonnentag06}.
As a result there is no single correct solution.
Therefore evaluations of programming and development proficiency must
contend with the need to evaluate the quality of the results.
Tests therefore sometimes focus on small well-defined tasks rather
than the general development process.

In the 1960s IBM devised the programmer aptitude test, which was
designed to aid in the evaluation of job candidates%
\footnote{J. L. Hughes and W. J. McNamara, IBM Programmer Aptitude
  Test (Revised), Form Number 120-6762-2.
  A scan is available at URL
  http://ed-thelen.org/comp-hist/IBM-ProgApti-120-6762-2.html.}.
This is actually more of a standard IQ test, with questions about
completing series of numbers, series of shapes, and high-school
arithmetic word problems; it has nothing to do directly with programming.
Nevertheless, this and some other tests became used by a large
percentage of the industry, and led to the formation of the ACM
special interest group on Computer Personnel Research (SIGCPR)
\cite{mayer68}.
However, the tests' ability to predict on-job programming performance
was debated.

Some ideas about assessing programming ability have been gleaned based
on analyses of the Advanced Placement Computer Science A tests
administered in the United States by the College Board, a private
company tasked with developing and administrating various standardized
tests.
This is a large-scale test offered to high-school students, meant to
be commensurate with the level achieved following a first-semester
college course in Java.

The first analysis of such test results was performed by Reges, who
analyzed the 1988 test results \cite{reges08}.
Interestingly, he found that five specific questions had non-trivial
correlations with all the rest, and therefore provide a good
indication for overall success.
(The question with the highest correlation was ``If \code{b} is a Boolean
variable, then the statement \code{b := (b = false)} has what
effect?'', and made it to the title of Reges's paper.)
It was speculated that these questions thus capture some core
elements of computer science understanding.
The same data was later re-analyzed by Lam et al., who concluded that
questions with legal code snippets in them were those that had the best
correlation with other questions (and hence with overall score)
\cite{lam12}.
However, this is tainted by the fact that nearly two thirds of the
questions were in this category.
They also identified questions about arrays, linked lists, compilers,
and recursion as having relatively high correlations.
Questions about invariants or including type definitions had
relatively strong negative correlations.

A more detailed analysis was conducted by Lewis et al.\ using the
results of the 2004 and 2008 tests \cite{lewis13}.
Note that the tests had of course changed since 1988.
For example, the programming language used was switched to Java.
More importantly, more questions included code in them, and by 2009
this applied to fully 95\% of the test.
Therefore having code could not be considered a discriminatory
characteristic of different questions.
This study failed to recognize any specific common
features in questions that are better at discriminating among test
participants, even though such questions were indeed identified.

In a related vein, Ma et al.\ studied the results of first year
students learning to program in Java.
They claim that a strong correlation exists between final grades
and holding valid mental models of value and reference assignments
\cite{ma07}.
Likewise, Simon et al.\ claim that the ability to clearly articulate a
problem-solving strategy is a good predictor of success in competence
tests \cite{simon06}.
These examples imply that the tests indeed measure deeper
understanding and not only superficial technical competence.
But Bayman and Mayer report a contrary finding, where many novice
programmers had misconceptions about what various BASIC statements do
despite being able to pass tests \cite{bayman83}.
Also, the old study by Sackman et al.\ found little correlation
between grades in tests and no-job performance \cite{sackman68}.

In the context of empirical software engineering research,
the idea is to use a separate common pretest task which is done by
all subjects to form a baseline for comparison \cite{arisholm04}.
Then different subjects do different treatments, and their results are
compared with their performance on the common initial task.
This has the additional beneficial attribute that the pretest measures
performance directly, and avoids all the confounding factors related
to experience, education, and motivation.

Pretesting has been used in several studies, but surprisingly
there has been very little work on devising the tests.
A specific test to measure programming skill has recently been
proposed and validated by Bergersen et al.\ \cite{bergersen14}.
This involves 12 Java programming tasks, and takes several hours, so
it is suitable for only large experiments.
Importantly, it combines scores for quality of the solution with the
time needed to achieve this solution \cite{bergersen11b}.

It has also been suggested that the pretest can focus on specific
relevant aspects of proficiency and knowledge
\cite{carver04,ko15,acuna06}.
But one should be aware of a dangerous loop.
In many cases, proficiency is included as one of the independent
variables that may affect and explain performance differences.
But if the test used to assess proficiency is similar in any way to
the experimental procedure being investigated, this may lead to a
meaningless finding that those who are better at performing the test
are also better at performing the related experiment.
For example, Bateson et al.\ set out to show that experienced
programmers (students who had taken more than 3 programming courses)
have better memory and problem solving skills than novices (students
who had taken up to 3 programming courses), by using tasks and
questions similar to those used in final exams of programming courses
\cite{bateson87}.
While they found that the experienced subjects did better, this
provides little information beyond the observation that students
retain at least some of what they had learned.
It is therefore necessary to devise general proficiency tests which
are independent of the issues being studied \cite{brooksre80}.

Siegmund et al.\ studied the use of background questions for assessing
programming-related proficiency.
To do this they considered the correlation between answers to various
potential questions and performance in 10 program comprehension tasks,
based on the assumption that more capable subjects should be able
to perform better \cite{siegmund14b}.
The result was that the highest correlation was obtained for the
question ``On a scale from 1 to 10, how do you estimate your
programming experience?'' (Spearman correlation coefficient of $\rho =
0.539$), and the second highest was for the question ``How do you
estimate your programming experience compared to your class mates?''
($\rho = 0.403$).
The question about years of programming experience led to a somewhat
lower yet statistically significant correlation ($\rho = 0.359$).
Questions about level of education (e.g.\ number of courses) led to
much lower and statistically insignificant correlation.

Kleinschmager and Hanenberg likewise compared the use of
pre-experiment tests, grades in university courses, and
self-assessment by the subjects themselves, in two separate
experiments involving 20 and 21 students.
In both cases the self-estimates allowed them to divide the students
into three groups, such that the performance of the top and bottom
groups were significantly different \cite{kleinschmager11}.
Course grades and pre-testing did not lead to such results;
for example, in one of the experiments students with lower grades in
programming courses actually performed better in the experiment.
Thus in their admittedly limited sample using grades or tests were not
better than relying on the self assessment of the subjects.
This may be explained by students studying together having a better
notion of how they really compare with each other.
But it is not clear that such self assessment would work equally well
in a more general setting.

A third related study, by Aranda et al., found that self assessment of
expertise was not correlated with effectiveness ($\rho$ between $-0.01$
and $-0.05$), and might suffer from over-estimation bias
\cite{aranda14}.
In this case experience fared slightly better, but again the sample
size was small.

The bottom line of this discussion is that
\mybox{It is hard to devise a good proficiency test, and more work on
  this is issue is required.
  For the time being, self-assessment appears to be a reasonably good
  option.}

\subsection{Limitations of Controlled Experiments}

In the preceding sections we discussed the meaning of being
a student and the need for tests which differentiate experimental
subjects according to their abilities.
Here we consider another dimension of the problem altogether: what
controlled experiments can be used for.
The idea is that if controlled experiments are in general of limited
scope, then maybe it is acceptable for the experimental subjects to be
limited too.

We have already noted above that controlled experiments in software
engineering are often chastised for using toy problems
(e.g.\ Figure \ref{fig:prob} and the related discussion).
The reason for using tasks with very limited scope is to keep the
experiments tractable, allowing subjects to complete them in a
relatively short time, and enabling the collection of results from
multiple subjects in the interest of statistical power.
But it is plausible that there are issues which simply cannot be
investigated using such controlled experiments, because they cannot be
condensed into a suitably limited framework.

One example for this is the effect and possible benefits of test-first
development.
Test-first development is a component of agile methodology, and in
particular of Extreme Programming (XP) \cite{beck99}.
A meta-study conducted by Bannerman and Martin showed that half
of the 6 controlled experiments investigating this issue using students 
found that test-first led to faster progress, and a solid majority
found that it had no effect on quality \cite{bannerman11}.
Conversely, of 11 studies using professionals, which were
predominantly case studies, 4 found that test first was slower, and 8
found that it led to higher quality.
Thus the results interacted not only with the experimental subjects,
but also with the study type, and perhaps the different results were
not due to using students but due to using controlled experiments.
In a related vein, di Bella et al.\ argue that studying pair
programming --- a basic component of XP --- should be done by
observing industrial developers in action over a prolonged period
rather than using small test cases in controlled experiments
\cite{dibella13}.

\hide{
Another example is afforded by the question of software evolution, and
specifically how projects grow with time.
In early work on evolution, Turski and Lehman suggested an
inverse-square model which implied that growth should be expected to
be sub-linear, meaning that the rate of growth will slow down as the
system grows and becomes more complex \cite{turski96}.
This was based on observations from large-scale industrial systems
such as OS/370.
But this model was challenged by later data from Linux, which
indicated that growth is super-linear and more specifically quadratic
\cite{godfrey00}.
The interpretation of such growth was that is was achieved by
employing a growing developer base \cite{capiluppi07b,feitelson12}.
Subsequent studies of Linux and other mainly open-source projects
provided mixed results, mostly exhibiting super-linear growth but
sometimes linear of even sub-linear growth
\cite{izurieta06,koch07,herraiz06,robles05,israeli10,feitelson12}.
Be that as it may, the obvious conclusion is that this issue has to be
investigated using observations, analyses of case studies, and perhaps
interviews with project managers.
It is outside the realm of controlled experiments.
}

More generally, it seems to be agreed that controlled experiments are
ill-suited to answer the ``big questions'', such as the benefits of
object-oriented programming and agile development, the patterns of
software evolution, and so on.
These issues encompass an extremely wide and complicated set of
factors and interactions, and really come into play mostly over
extended periods and large projects.
This realization has led to the growth of the whole field of mining
software repositories as part of performing case study research, often
augmented by surveys of professionals, and leading to theory building.

Furthermore, controlled experiments can also not be used when studying
professional work practices and the dynamics of large-scale software
development, e.g.\ the communication between different teams taking
part in a project \cite{damian13}.
Thus
\mybox{Assuming we accept that controlled experiments are limited to
  relatively small, focused, and well-defined tasks, the expertise of
  experimental subjects may be of limited importance and students may
  be suitable.}

\subsection{Other Issues and Confounding Factors}

The obsession with the question of using students as experimental
subjects may mask other important factors.
First, as we have discussed extensively, it may actually be the wrong
question, because what we are really interested in is probably
proficiency.
Second, there are other potentially important factors that receive
much less attention, such as sex, personality, and mood
\cite{hannay11,graziotin14}.

In describing empirical studies one often includes demographic data
such as the number of male and female subjects, but most often this is
not used as an independent variable when analyzing the results.
One exception is the work on program reading patterns by Sharafi et
al., in the context of a study on different identifier styles.
Specifically, they write that ``male and female subjects
follow different comprehension strategies: female subjects seem to
carefully weight all options and spend more time to rule out wrong
answers while male subjects seem to quickly set their minds on some
answers, possibly the wrong ones'' \cite{sharafi12}.

Unlike age and sex, the personality of experimental subjects is
typically not assessed in the context of controlled experiments.
This does not mean that it is not important.
Personality can be characterized in various ways.
One of them is the Myers-Briggs personality type notation, which
consists of a combination of 4 letters \cite{devitodc07}:
\begin{description}\itemsep 0pt
  \item [E/I] denotes being and \textbf{e}xtrovert or
    \textbf{i}ntrovert.
    Extroverts thrive on human interaction, whereas introverts tend to
    be loners.
  \item [S/N] denotes relying on the \textbf{s}enses to obtain
    information, or alternatively relying more on i\textbf{n}tuition.
  \item [T/F] denotes the way decisions are reached: either
    \textbf{t}hinking it out logically, or based on \textbf{f}eelings
    and the opinions of others.
  \item [J/P] denotes using \textbf{j}udgment to plan everything in
    advance, as opposed to using \textbf{p}erception and being more
    spontaneous and flexible.
\end{description}
Such classifications have been used to classify programmers and check
whether their personality affects their performance.
For example, Capretz studied a sample of 100 software engineers and
found that a full 24\% of them were of ISTJ type, more than double the
rate in the general population \cite{capretz03}.
The most significant dimension was T/F, with 81\% being thinking
to only 19\% feeling.
Followup work suggested that different personality types are suitable
for different roles and different stages in the software lifecycle
\cite{capretz10}.
Conversely, Turley and Bieman found that among exceptional software
engineers (albeit a small sample) half were INTJ, that is, using
intuition rather than sensing information, and 85\% overall were
thinking types \cite{turley95}.
Bishop-Clark and Wheeler conducted a study showing that
sensing (S) students performed better than intuitive (N) students, and
judging (J) students were better than perceptive (P) students,
but only on programming tasks and not in final grades
\cite{bishopc94}.
Similar studies have implicated personality types in the success of
pair programming \cite{begel08,walle09,salleh09} and in code reviews
\cite{devitodc07}.
Finally, Turley and Bieman claim that when trying to classify software
engineers into exceptional and nonexceptional ones, general
personality traits such as ``helps others'' and ``willing to confront
others'' were among the dominant discriminant factors \cite{turley95}.

\mybox{Personality and sex are attributes of experimental subjects
  which may be just as important as expertise.}

The subjects being used are an important attribute of controlled
experiments, but there are others \cite{dyba12}.
According to Votta and Porter, empirical research in software
engineering must contend with 3 dimensions \cite{votta95}:
\begin{itemize}\itemsep 0pt
\item Individuals vs.\ groups
\item Students vs.\ professionals
\item Lab conditions vs.\ real life
\end{itemize}
We have focused on the second of these, and extended the discussion to
also include different levels of proficiency.
But this may not be the most important dimension.
Controlled experiments nearly always use subjects working
individually, and in laboratory conditions, namely tackling
well-defined problems of modest size.
This may be cause for significant threats to external validity.

Working in groups may be different.
First there is the issue of division of labor, where each developer
needs to focus on only part of the problem.
This also facilitates a measure of specialization, allowing different
developers to become experts in their individual areas of
responsibility.
Second, group dynamics and interactions facilitate cross
fertilization and the exchange of ideas, ideally leading to a
situation where the whole group performs better than the sum of its
individual participants.
Thus the social context has an effect on how people work \cite{riedl91}.
Conversely, personality clashes may come into effect in groups and
disrupt their progress, an effect that would not occur when developers
work individually.
To account for such effects, Basili and Zelkowitz suggest to perform a
series of experiments progressing from individual programmers to
groups, and from students to industrial professionals \cite{basili07}.

Likewise, there are possible interactions with problem size.
The toy systems used in experiments may be unrepresentative of real
problems encountered in professional work.
In particular, it has been observed that real complex problems are
needed to bring out the advantage of expertise \cite{schenk98}.
Putting these two aspects together, Damian et al.\ report on
research showing that team members with different roles and different
domain knowledge hold the key to effective functional communication in
a project \cite{damian13}.
This study naturally relied on observations, interviews, and surveys
of professional practitioners involved in multi-team projects, and
could not be done with students in a lab.

However, one should also not exaggerate the importance of working
in groups or on large and complex problems.
The importance of these factors may depend on the study.
When studying code comprehension, for example, individuals and limited
code may well suffice.
When studying tool usage individuals again may suffice, but the amount
of code needed may be larger in order to bring out the benefits of
using the tool.
Then there is the question of whether we are seeking relative results
(a comparison of two methodologies to see which is better) or more
precise absolute ones.

\mybox{The study context, e.g.\ working in a group or in real-life
  conditions, may perhaps be more important than attributes of
  experimental subjects.}

\subsection{Compensation by Experimental Procedures}

The main perceived problem with using students, as articulated in
many of the works cited above, is that they are inexperienced and
therefore not representative.
Another problem is the danger that there will be a very wide range of
results, and that this variability would swamp out the experimental
effects \cite{curtis14}.
Therefore such variability needs to be controlled or at least factored
out, by using appropriate experimental methodology.

An interesting approach to reduce the adverse effects of conducting
experiments with inexperienced students was suggested by
Carver et al.\ \cite{carver03b}.
The concern was that when evaluating a new technology the students are
too low on the learning curve, so their achievements will not be
representative of professionals who spend more time to learn the new
approach.
The suggested solution was to divide the students into pairs, and
conduct the study twice: first one performs the task and the other
observes, and then they switch roles and do it again.
The measurement is done only on the second time.
The first is used only to accelerate the learning by providing
hands-on experience.

Another interesting alternative for assessing the effect of
experience is to do so in reverse.
Bednarik and Tukiainen identified two strategies of performing
comprehension tasks in their eye-tracking study \cite{bednarik06}.
So they retroactively divided the study participants according to the
strategy (in particular, how many program animation runs they used),
and analyzed the background data on the two groups that were produced.
The results were that participants in one group had significantly more
previous programming experience than participants in the other group,
and also included the two professional programmers that had
participated in the study.

Regarding the effect of the large variability of results, using
students may actually reduce the variability because they all have
about the same level of education, leading to better statistical
characteristics \cite{basili99,ko15}.
Nevertheless, one should always apply an independent test
of general proficiency and filter outliers to reduce variability.
Alternatively, it may be possible to create teams with the same mix of
proficiencies, e.g.\ by identifying the top-performing students and
distributing them among the teams \cite{basili96b}.

Brooks suggests the use of within-subject experimental designs to
compensate for differences in ability \cite{brooksre80}.
Variance is reduced and the analysis is indeed improved by
analyzing the distribution of within-subject differences between the
treatments, rather than trying to first characterize the performance
for each treatment individually \cite[chap.\ 11]{law:book}.
Within subject designs are a special case of block designs, where
levels of the uninteresting factor (in our case, the experience or
proficiency of experimental subjects) are randomly balanced across
experimental treatments.
This and other approaches to statistical analysis are discussed at
length by Juristo and Moreno \cite{juristo:exp}.
A problem with such analyses is that traditionally the analysis was
based on assumptions relating to the normality of the data.
However, robust alternatives that apply to any distribution are also
available \cite{wilcox:book1,wilcox:book2}.

The advantage of within-subject designs stems from the fact that each
subject is exposed to all the different levels of the treatment.
Thus, if a certain subject's abilities are either above or below the
norm, this will apply to all the treatments in the same way.
However, if the task being performed is the same, there is a
significant risk of a learning effect \cite{kitchenham03}.
In the first instance the subject needs to contend both with a new
task and with the specific treatment being used, but in the second
instance the task is already known, so only effects of the treatment
remain.
Using different tasks eliminates the learning effect, but introduces
the task as a new (and possibly no better) confounding factor.

\mybox{It is desirable to focus on within-subject differences
  when mixed populations of subjects are used, provided learning
  effects can be ruled out.}

\hide{
The use of statistics is needed when a number of experimental subjects
are used, and indeed it is commonly thought that the more subjects the
better.
A completely different approach was suggested by Harrison (but it is
not clear if it was ever used in this context) \cite{harrison97}.
This is the notion of using $N=1$, namely a single experimental
subject instead of a large sample.
The claim was that if we focus on a single subject, we can get the
following benefits:
\begin{itemize}\itemsep 0pt
\item Use a professional developer working on real problems in a real
  work environment, neutralizing the danger of non-representative
  results.
\item Conduct the study over extended periods of many months, allowing
  subtle effects and effects that take a long time to materialize to
  be observed.
\end{itemize}
It should be noted that such single-subject experiments are still
experiments, and not case studies.
The crux is in using an A-B design (or an A-B-A design) where a single
factor is changed in a controlled manner during the observation period
to assess its impact.
}

Finally, an important experimental tool is replication.
In particular, external replication (that is, replication by other
researchers) is considered an especially effective technique to
increase confidence in a result.
This increase in confidence is contingent on the unavoidable
variations between the original and the replication, which show that
the observed effect is indeed robust.
In the context of software engineering experiments, a major element of
this variability is the use of different experimental subjects
\cite{shull08,juristo11,feitelson15}.
And in particular, replications using professionals can increase the
confidence in student-based results.

\subsection{Students Explicitly Desired}

Nearing the end of this discussion, we consider a special case of
software engineering experimentation which explicitly targets novices,
e.g.\ to see how certain tools may help them to perform certain tasks
despite lack of prior knowledge or experience \cite{kitchenham02}.
In such situations is may actually be appropriate to focus on
undergraduate students and exclude more mature students in order to
reduce the variability in the subjects' level of experience.

For example, Liu et al.\ describe a tool called InsRefactor which is
designed to help novice programmers refactor their code and resolve
code smells \cite{liu13}.
The idea is to proactively alert them to code smells as they are
created, rather than leaving it up to them to request information
about code smells retroactively.
To investigate the effectiveness of this tool a controlled experiment
with two groups of students was used.
Similarly, Fernandez et al.\ used students to test a usability
inspection tool integrated into a web development process, and
explicitly justify this by citing ``the intention ... to provide a Web
usability evaluation method which enables inexperienced evaluators to
perform their own usability evaluations'' \cite{fernandez13}.
A third example is the work of van Heesch et al., who studied the
degree to which documentation helps junior designers
\cite{vanheesch13}.
And Busjahn et al.\ use novice students specifically as a
contrast to experienced professionals, to show how experience affects
code reading patterns \cite{busjahn15}.

Briand et al.\ conducted a study of how quality guidelines affect
maintainability, using students as subjects \cite{briand97,briand01}.
In particular, they identify the cognitive complexity of
object-oriented designs as a potential problem.
While they frankly note that this causes a threat to external validity as
it is not clear that the results generalize beyond inexperienced
students, they argue that inexperienced programmers are often assigned
to maintenance tasks, and therefore student subjects are actually
appropriate in this case.

Another situation where students are explicitly needed is in the
evaluation of educational tools and procedures.
One example is the study by Runeson concerning the effectiveness of
PSP (Personal Software Process) training \cite{runeson03}.
He found that the improvements from PSP level 0 to PSP level 2 was
similar for freshmen and for graduate students.
Another example is described by Janzen et al.\ \cite{janzen13}.
This is a study of an educational platform called WebIDE which
was used to teach introductory Java and Android programming.
A controlled experiment was used to compare a group of students who
used this platform with another group who used a more conventional
lab setting.
A third example is a study on the effect of internship on on-job
performance \cite{dossantos13}.
In this case two groups of interns were assessed for five months as
they used agile methods on projects in the telecom industry.

Finally, Kuzniarz et al.\ suggest that the worst-case experiment for a
new methodology or tool is when subjects know about the (established)
experimental alternative but not about the new treatment
\cite{kuzniarz03}.
Under this scenario a positive effect regarding the experimental
treatment is especially convincing.
And students are especially suitable for such experiments, as their
lack of experience increases the chance that they do not know of the
new treatment.

\section{Recommendations Regarding Using Students}
\label{sect:rec}

As expanded in Section \ref{sect:conc} below, our main conclusion is
that the issue of using students as experimental subjects should not
be the factor which determines whether an empirical study is
considered worthwhile.
Students may be used beneficially in controlled experiments in many
cases, and as far as their availability leads to conducting more such
experiments their use may catalyze significant contributions to
software engineering research.
But there are indeed cases where students would be inappropriate.

The main consideration regarding the use of students is that their
level should be matched to the requirements of the study being
performed.
This leads to the following more specific recommendations:
\begin{enumerate}\itemsep 0pt
\item Studies of problems in beginning to program, programming
  education, or non-programmer end-user assessment can use novice
  students (in their first year, after one programming course).
\item Studies that do not require an extensive learning curve can use
  intermediate students (toward the end of their BSc, but with no
  industrial experience).
  In effect such students can be expected to have similar capabilities
  as beginning professional developers, leading to valid relative
  results when comparing straightforward tools or methodologies.
\item More precise quantitative studies or those requiring more
  experience may use advanced students (graduate students in a
  programming-related program and/or with industrial experience).
  However, it is always advisable to assess them individually and not
  count on their schooling and experience alone --- and this applies
  also to non-student subjects.
  Training should be provided as appropriate.
\end{enumerate}
Students should generally \emph{not} be used in studies that depend on
specific expertise which requires significant experience and a long
learning curve to achieve, or in studies of professional practices.
Such studies are best performed by observing and interviewing
professionals, not by controlled experiments.

All the above is from the scientific and experimental point of view.
But using students also has ethical aspects.
Ethical concerns usually relate to avoiding inflicting any harm on the
experimental subject or on society at large, and on informed and
voluntary consent to participate in the experiment.
This is regulated by Institutional Review Boards (IRBs)
\cite{king15}.

In the context of student participation in software engineering
experiments there is no real danger of causing actual harm to the
students, but some have voiced concern regarding harm to their
academic progress.
Therefore, especially when experiments are carried out as part of
compulsory classes, they should have educational goals
\cite{carver03,berander04}.
Examples include the opportunity to learn or exercise some technique or
methodology,
being exposed to cutting-edge ideas and procedures,
creating awareness of difficulties and trade-offs,
providing industrial-like experience,
and first-hand learning about empirical methods.
Evidence that participation in experiments indeed contributes to
students' education has been provided by Staron \cite{staron07}.

Still, using students in experiments should be done subject to ethical
considerations \cite{singer02,carver03,carver10}.
For example, could the students have learned the same things more
efficiently in some other way?
Is it fair and reasonable to grade them on their performance in an
experiment, especially if they were divided into groups that used
different treatments?
Is it reasonable to give academic credit for participation in experiments?
Some of these concerns can be mitigated by giving students feedback
after the experiment is completed, so they see how their participation
added to the knowledge in the field.

The issue of informed consent is also problematic in a classroom
setting, as students may refrain from opposing suggestions or requests
from their professors \cite{singer02}.
Thus at a minimum one needs to uphold anonymity, and allow the option
to opt out, thereby negating the fear of influence on grades.
Thus the ethical point of view in this matter coincides with the
methodological view that coercion to participate in an experiment may
lead to unreliable results --- a problem that can occur also in an
industrial setting, and is not unique to academia.
Indeed, it is in general necessary not to mix experimental
observations with evaluations of performance.
Gathered data should not be used to evaluate subjects outside of the
study context.

\section{Conclusions}
\label{sect:conc}

Using students in software engineering experiments is often cited as a
problem, because students constitute a convenience sample: they are
selected for the study because they are easily available to the
researcher, and because they are cheap, regardless of whether they are
representative of the target population in general.
This is a far cry from the ideal of using random sampling, where study
participants can be argued to truly represent the target population.
As a result many researchers (and reviewers) have reservations about
the external validity of student-based experiments, claiming there is
no reason to believe that the results generalize beyond the original
study population.

Conversely, it has been claimed that in many contexts using student
subjects is actually valid.
While the students are not a valid representative statistical sample
of software professionals, they can be viewed as the next generation
of professionals \cite{kitchenham02,tichy00}.
So students are perfectly suitable when the study does not require a
steep learning curve for using new technology \cite{basili99,salman15}.
Moreover, using industry professionals or web-based volunteers is
usually not any better, because these techniques too cannot guarantee
a valid random sample of the general software practitioner population.
And in any case there are also wide differences in background,
experience, and capabilities among practitioners.

The main conclusions of this review are summarized in Table
\ref{tab:conc}.
Taken together, the overarching message is that ``can students be used
as experimental subjects?'' is not the right question.
First and foremost, the goal should be that 
\emph{the observed effect be representative of the real effect},
rather than that the experimental subjects be representative of real
developers.
But even this is an over-simplification, because it assumes that there
is a single real effect.
As a research community, we need to embrace variability and collect
much more data from diverse conditions.

\begin{table}\centering
  \caption{\label{tab:conc}\sl
    Summary of main observations and recommendations.}
  \begin{tabular}{rp{0.88\textwidth}}
    \hline\\
1.& The level of the experimental subjects (students or otherwise)
    should be commensurate with the tasks they are expected to
    perform\\[3mm]
 
2.& Graduate students and even students at the end of their BSc
    have similar proficiency to industrial professionals for general
    programming tasks\\[3mm]

3.& In many cases experiments with student subjects lead to the
    correct relative results, even if they are generally
    slower or less proficient than professionals\\[3mm]

4.& Students are naturally suitable when studying novices\\[3mm]

5.& Subjects with experience with the tools or techniques being
    studied are required when there is a long learning curve to use
    them effectively; in some cases adequate training sessions should
    be provided\\[3mm]

6.& Studies of how experts tackle complex problems require real
    experts, but also real problems; performing them is especially
    difficult, and alternatives such as case studies should be
    considered\\[3mm]

7.& When the effect of proficiency is the focus of study, subjects
    should be assessed individually rather than being assigned by
    degree or affiliation\\[3mm]

8.& Academic studies with students can be profitably used in initial
    steps of larger industrial collaboration research efforts\\[3mm]

9.& One can compensate for different levels of proficiency by using
    within-subject experimental designs, provided there is no danger
    of learning effects\\[3mm]

10.& Subjects should want to participate in the experiment.
    Students or professionals who are told to participate may be
    problematic.\\[3mm]

11.& Experience is just one factor that may affect proficiency, and it
    may be important to consider others as well (e.g.\ sex or
    personality)\\[3mm]

12.& When using students as experimental subjects,
    \begin{itemize}\itemsep 0pt
    \item Student subjects should not be identified and their performance in
      the experiment should not affect their grades
    \item Participation in experiments should not be compulsory for students
      unless the experiment has clear educational benefits
    \end{itemize}\\
    \hline
  \end{tabular}
  
\end{table}

Reviewing the literature on the subject indicates that ``students
vs.\ professionals'' is actually a misrepresentation of the
confounding effect of proficiency, and in fact differences in
performance are much more important than differences in status
\cite{sonnentag06,soh12}.
It is reasonable to assume that there indeed exists a big difference
between complete novices and graduating students, but after three or
four semesters students are already reasonable experimental subjects
for general studies.
It is possible that upon starting employment there is another
large increase in capabilities, but this may be more focused on the
specific technologies used in the individual place of work.
Indeed, a major problem in experimental software engineering is the
differences between individual experimental subjects.
Such differences suggest the need for a basic proficiency test as part
of the experimental setup \cite{feigenspan12,siegmund15}.
But it is not easy to come up with a simple and discriminating test.

Given the difficulty in measuring proficiency, experience is often used
as a proxy, under the assumption that more experience is equivalent to
higher proficiency.
This may be true for relative beginners (students and new
professionals), but more senior professionals may reach a plateau
where additional experience does not lead to significant additional
improvements \cite{sonnentag98}.
Still, it is important to acknowledge that expertise (typically
emanating from experience) does in fact lead to different behaviors in
some cases.
Thus professionals (or experienced developers) are needed when
maturity such as in tool usage is potentially part of the setup.
Inexperienced students don't use tools as much, even if they could
benefit from them more \cite{ceccato14}.

Special care should be taken when claims about interactions with
experience or proficiency are made.
One possible problem is confounding proficiency-based filtering with
the results, such that subjects identified as less proficient will
necessarily perform worse.
Another is that claimed results about the effect of experience,
e.g.\ that a tool helps inexperienced developers more, may not be
convincing because not enough experienced developers were available to
compare with, and it may be the case that the tool helps for everyone
\cite{roehm13}.

The alternative to using students is to recruit industrial
professionals.
This may be hard to do: one either has to pay them a competitive fee,
or at least schedule experiments so as not to interfere with their
normal work schedule \cite{sjoberg02,dieste13,vegas15}.
Therefore it is important to make the best use of this scarce
resource.
One repeated suggestion is to conduct a pilot study first, based on
students, and only then move to the ``real'' study with professionals
\cite{sjoberg03,carver03,gorschek06,basili07}.
Using students first allows to both test and debug the experimental
procedure, and to justify the extra effort of using professionals
\cite{basili99,carver10,tichy00}.

Another issue is where the experiment is run: maybe conducting it in
an industrial setting is even more important than using professionals
\cite{sjoberg02,dyba12,dieste13}.
But companies are often reluctant to support this for lack of
immediate tangible benefits.
This is short sighted, as industry can also benefit from participating
in studies \cite{carver03}.
Specifically, by allowing employees to participate in academic
experiments they obtain early evidence to confirm or refute hypotheses
about methodologies and technologies; they learn about new ones; they
obtain knowledge of the procedures, costs, and benefits of empirical
software engineering; and they learn about their own process and
people (in observational/survey studies).
One can also view experiments as training \cite{laitenberger97} or
technology transfer \cite{linkman97,basili93,basili07,gorschek06}, and
possibly use internship in the company as a vehicle \cite{causevic13},
which may later aid in recruitment.

It is important to note that the issue of using students is but one
aspect of empirical software engineering studies.
There are many other methodological issues that are no less important.
One issue that is often problematic is using appropriate statistical
tools.
This starts with the most basic, for example the need to show full
data distributions, or at least box plots, and not make do with
averages of widely dispersed data points with skewed or bimodal
distributions.
It continues with using modern and non-parametric techniques that are
not compromised by data which does not fit model assumptions like
normality \cite{wilcox:book1,wilcox:book2}.

There is also significant need for more experimentation in general
\cite{tichy95,tichy98}.
A literature study of the period 1993--2002 found that controlled
experiments papers represent only 1.9\% of all 4543 papers that were
examined \cite{sjoberg05}.
Indeed, too many results in the literature, which are then
accepted as ``facts'', are actually based on a single experiment in a
specific setting with few subjects conducted many years ago.
A selection of interesting examples is provided by Glass
\cite{glass:facts}, including the often-quoted 28:1 ratio in
performance between the best and worse programmers, which is actually
based on a study of debugging performance from 1968 using a grand
total of 21 subjects \cite{sackman68}.

Discussions and arguments for external validity of individual studies
cannot replace actual experimental evidence.
In particular, it is ludicrous to expect any single study to reveal the
full picture or even a significant part of the picture of a research
topic.
Rather, each individual study should be regarded as a pixel.
Moreover, at least some of them should be targeted at basic science
issues with no immediate practical relevance.
The full picture can then emerge when we have enough replications of
enough diverse studies --- orders of magnitude more than we have now
---  to step back and observe the underlying currents and revealed
patterns.

To obtain the required evidence many more replications are needed, as
opposed to branching into innovative uncharted territory.
Moreover, different levels of replications should be employed
\cite{feitelson15,juristo11,shull08}.
This includes variations in the experimental artifacts and the
approach in addition to using different experimental subjects.
Experimental validity can only be obtained by using all of these in
tandem.

And finally returning to the issue of students as experimental
subjects, our understanding of using different experimental 
subjects will improve with more empirical work on the effects of
experience and expertise, and specifically on what makes
highly-performing developers different and how expertise develops or
can be promoted \cite{ericsson93,sonnentag06,dreyfus80,schenk98}.
This line of work is more meaningful than comparing students to
professionals.

\subsection*{Acknowledgments}
This research was supported by the ISRAEL SCIENCE FOUNDATION (grant
no.\ 407/13).
Many thanks to Lutz Prechelt for his very useful comments on an
earlier draft of this paper, and to anonymous reviewers of another
draft.

\bibliographystyle{myabbrv}
\bibliography{abbrv,par,misc,se}
\end{document}